\documentclass[journal]{IEEEtran}

\usepackage[utf8]{inputenc}
\usepackage{amsmath}
\usepackage{cite}
\usepackage{multicol}
\usepackage{graphicx}
\usepackage{array}
\usepackage{multirow}
\usepackage{graphicx}
\usepackage{enumitem}
\usepackage{color}
\usepackage{nomencl}
\usepackage{etoolbox}
\usepackage{amssymb}
\usepackage{comment}
\usepackage{mathtools}
\usepackage{bbold}
\usepackage{yfonts}
\usepackage{soul}
\usepackage{hyperref}
\usepackage{dsfont}
\usepackage[acronym]{glossaries}
\usepackage{fancyhdr}

\usepackage{fnpct}
\ifCLASSOPTIONcompsoc \usepackage[caption=false,font=normalsize,labelfon
t=sf,textfont=sf]{subfig}
\else
\usepackage[caption=false,font=footnotesize]{subfig}
\fi
\usepackage{tikz}
\usetikzlibrary{decorations.pathreplacing}
\usetikzlibrary{arrows,shapes,positioning}
\usetikzlibrary{patterns}
\usepackage{pgfplots}
\pgfplotsset{compat=newest}
\pgfplotsset{plot coordinates/math parser=false}
\newlength\figureheight
\newlength\matlabfigurewidth
\newcommand\Item[1][]{%
  \ifx\relax#1\relax  \item \else \item[#1] \fi
  \abovedisplayskip=0pt\abovedisplayshortskip=0pt~\vspace*{-\baselineskip}}

\title{Grid-aware Distributed Model Predictive Control of Heterogeneous Resources in a Distribution Network: Theory and Experimental Validation}
\author{
    \IEEEauthorblockN{
    Rahul Gupta\thanks{$^*$Corresponding author: rahul.gupta@epfl.ch.}$^{*,1}$, Fabrizio Sossan$^{2}$,  Mario Paolone${^1}$\\}
    \IEEEauthorblockA{$^{1}$Distributed Electrical Systems Laboratory, EPFL, Switzerland,
    $^{2}$PERSEE, Mines ParisTech -- PSL, France
   }}
\usepackage{etoolbox}
\makeatletter
\patchcmd{\@maketitle}
  {\addvspace{0.5\baselineskip}\egroup}
  {\addvspace{-2\baselineskip}\egroup}
  {}
  {}
  
\begin{document}
\markboth{IEEE Trans. On Energy Conversion, to be published (2020).}%
{}
\maketitle

\begin{abstract}
In this paper, we propose and experimentally validate a scheduling and control framework for distributed energy resources (DERs) that achieves to track a day-ahead dispatch plan 
of a distribution network hosting controllable and stochastic heterogeneous resources while respecting the local grid constraints on nodal voltages and lines ampacities. The framework consists of two algorithmic layers. In the first one (day-ahead scheduling), we determine an aggregated dispatch plan. In the second layer 
(real-time control), a distributed model predictive control (MPC) determines the active and reactive power set-points of the DERs so that their aggregated contribution tracks the dispatch plan while obeying to DERs operational constraints as well as the grid’s ones.
The proposed framework is experimentally validated on a real-scale microgrid that reproduces the network specifications of the CIGRE microgrid benchmark system. 
\end{abstract}

\section*{Acronyms}
\begin{table}[h]
    \centering
    \normalsize
    \renewcommand{\arraystretch}{1}
    {
    \begin{tabular}{l l}
         {ADMM} & Alternating Direction Method of Multipliers\\
         {BESS} & Battery Energy Storage System\\
         {DERs} & Distributed Energy Resources\\
        {DSO} & Distribution System Operator\\
        {GCP} & Grid Connection Point\\
        {GHI} & Global Horizontal Irradiance\\
        {GPS} & Global Positioning System\\
        {MAE} & Maximum Absolute Error\\
        {MPC} & Model Predictive Control\\
        {OPF} & Optimal Power Flow\\
        {PMU} & Phasor Measurement Units\\
        {PV} & PhotoVoltaics\\
        {RMSE} & Root Mean Square Error\\
        {SC} & Sensitivity Coefficient\\
        {SCADA} & Supervisory Control and Data Acquisition\\ 
        {SD} & Standard Deviation\\
        {SOC} & State-of-Charge\\
        {TSO} & Transmission System Operator\\
        {UDP} & User Datagram Protocol\\
        {UTC} & Universal Time Coordinated. \\
    \end{tabular}}
    \label{tab:my_label}
\end{table}

\section{Introduction}
Controlling heterogeneous and stochastic energy resources connected to medium and low voltage power grids is crucial to displace centralized electricity generation in favor of renewables.
This change of paradigm inputs the planning and operational practices of both distribution system operators (DSOs) and transmission system operators (TSOs). Indeed, DSOs may face significant issues associated with grid reinforcements and capability of being dispatched, while TSOs will experience increasing needs for allocating and deploying regulating power.

Day-ahead and intra-day scheduling of distributed energy resources (DERs) and, in general, heterogeneous DERs has been advocated in the literature as a way to minimize the effect of uncertainties. It consists in determining an average power trajectory (dispatch plan) at a certain resolution before operations that is then followed during real-time operation. 
Different scheduling problems have been proposed. E.g., works in \cite{6417004_short, 6913566_short, he2011multiple, sossan2016achieving_short} aim at minimizing power imbalances and in \cite{lorca2014adaptive} at maximizing economic benefit.
During operation, the realized power profile deviates from the dispatch plan because of forecast errors causing issues such as: power imbalance, lines/transformers congestion, voltage outside bounds etc. 
To tackle these issues, several works proposed real-time controls and energy management schemes 
\cite{Christakou_voltage_short, Bernstein2015_short, gupta2018admm_short, namor2018control_short} with different objectives. 
The work in \cite{Christakou_voltage_short} proposed a real-time control for voltage regulation, in \cite{Bernstein2015_short} for congestion management, in \cite{gupta2018admm_short} for energy management and dispatch tracking, and in \cite{namor2018control_short} for frequency regulation. An extensive literature review on micro-grids controls and energy management schemes is presented in \cite{espin2020energy_short}.

In this work, we develop and experimentally validate a framework for scheduling and real-time control of heterogeneous DERs while accounting for local grid constraints. In the scheduling phase on the day before operations, a stochastic optimization problem computes an aggregated dispatch plan at the grid connection point (GCP), accounting for the uncertainties of DERs and demand with scenarios, and constraints of the grid and DERs with models. %
In the real-time phase, a grid-aware model predictive control (MPC) computes the active and reactive power set-points of the heterogeneous DERs so that their aggregated contribution tracks the dispatch plan at the GCP while obeying to their constraints and those of the grid. The MPC problem leverages a distributed formulation to achieve a privacy-preserving and scalable configuration. It is solved with the alternating direction method of multipliers (ADMM), in which each DER solves its own constrained optimization problem, and an aggregator performs an optimal power flow (OPF) to enforce grid constraints and track the dispatch plan, considering grid losses. We model the grid with a linearized model based on the sensitivity coefficients \cite{christakou2013efficient_short, gupta2019performance} to retain the convexity and tractability of the underlying optimization problem and achieve the high computational performance necessary for the hard real-time control.

The proposed framework is experimentally validated in a grid-connected microgrid that reproduces the CIGRE benchmark system for microgrids \cite{papathanassiou2005benchmark}. It connects stochastic demand, a battery energy storage system (BESS) and two curtailable PV power plants and is equipped with a phasor measurement units (PMU)-based monitoring system.  We perform experiments for two distinct days characterized by different irradiance patterns, including real-life forecasters, also described in the paper. 

The main contributions of this paper compared to the existing literature are the formulation of a generic and computationally-efficient scheduling and control framework to dispatch heterogeneous resources while accounting for grid constraints and its experimental validation in a real-life setup. With respect to previous efforts of experimental validation in \cite{6417004_short, sossan2016achieving_short, gupta2018admm_short, namor2018control_short} and MPC-based control in \cite{perez2012predictive_short, halamay2014improving_short, teleke2009control_short, zheng2017distributed_short, du2017distributed_short}, we report the first experimental validation of a rigorous distributed MPC-based framework on a real-scale microgrid accounting for the grid constraints.

The paper is organized as follows. Section~\ref{sec:Prelim+Prob_stat} states the problem, Section~\ref{sec:Day_ahead} describes the day-ahead problem, Section~\ref{sec:Real-time} presents the real-time controller, Section~\ref{sec:Expt.} presents the experimental setup, Section~\ref{sec:results_discussion} presents the experimental results and Section~\ref{sec:conclusion} summarizes the outcomes and findings of the paper. 
\section{Problem statement}
\label{sec:Prelim+Prob_stat}
We consider a distribution grid interfacing controllable and uncontrollable DERs. The grid is dispatched at its GCP according to a pre-determined dispatch plan. The dispatch action is achieved by coordinating the DERs operations while respecting their own constraints along with those of the grid. 
We refer to the following two-stage mechanism.
\begin{itemize}[leftmargin=*]
    \item \textbf{Day-ahead scheduling:} the operator computes a dispatch plan for the next day based on the forecast of the stochastic generation and demand, the status of controllable resources, and local grid constraints. 
    The dispatch plan not only reflects the point-predictions of the stochastic quantities but also ensures that DERs have a suitable level of flexibility to track the dispatch plan in real-time. We assume that the dispatch plan has a 30-sec resolution and is computed at 23.00 UTC the day before operations. This phase is detailed in Sec.~\ref{sec:Day_ahead}.
    \item \textbf{Real-time operations:} DERs are controlled in real-time, so to compensate for power mismatches at the GCP between the realization and dispatch plan. The control problem is formulated as a distributed MPC. It accounts for future uncertainties along the optimization horizon, and DERs' and grid's constraints. The distributed formulation decouples the DERs' and the grid's problems, which can be solved iteratively until convergence.  Real-time operations start at 00.00 UTC and end at 23.59.59 UTC. The formulation is detailed in Sec. \ref{sec:Real-time}.
\end{itemize} 

In both stages, we model the grid with sensitivity coefficients (SCs), which express the nodal voltages, line currents, and grid losses as a linearized function of the nodal complex power injections. Let us consider a generic distribution grid with $n_b$ nodes and $n_l$ branches. Vectors $\mathbf{v} \in \mathds{R}^{(n_b-1) }$ and $\mathbf{i} \in \mathds{R}^{n_l}$ represent direct sequence nodal voltages magnitudes and branch currents magnitudes, respectively, and $\mathbf{p} \in \mathds{R}^{(n_b-1) }$ and $\mathbf{q} \in \mathds{R}^{(n_b-1) }$ the three-phase total nodal active and reactive controllable injections for all nodes except the slack node. Scalars $p^l, q^l \in  \mathds{R}$ are the total active and reactive transmission losses as seen at the GCP. The linearized nodal voltages, branch currents and the losses seen at the GCP are\footnote{We assume the following hypothesis i) the system is in steady-state and can be modeled by phasors, which is able to track small power-dynamics, ii) the nodes are PQ nodes, and the nodal injections are not voltage-dependent.}:
\begin{align}
\mathbf{v}_t&=\mathbf{A}_t^\mathbf{v} \begin{bmatrix}
    \mathbf{p}_t~
    \mathbf{q}_t
\end{bmatrix}^T 
+ \mathbf{b}_t^\mathbf{v}\label{eq:lin_grid_model_v}\\
\mathbf{i}_t&=\mathbf{A}_t^{\mathbf{i}} \begin{bmatrix}
    \mathbf{p}_t~
    \mathbf{q}_t
\end{bmatrix}^T 
+ \mathbf{b}_t^\mathbf{i}\label{eq:lin_grid_model_i}\\
\begin{bmatrix}
    p_t^l~
    q_t^l
\end{bmatrix}^T&=\mathbf{A}_t^l \begin{bmatrix}
    \mathbf{p}_t~
    \mathbf{q}_t
\end{bmatrix}^T 
+ \mathbf{b}_t^{l},
\label{eq:lin_grid_model_pl}
\end{align}
where $\mathbf{A}_t^\mathbf{v} \in \mathds{R}^{(n_b-1)\times 2(n_b-1)}$, $\mathbf{A}_t^{\mathbf{i}} \in \mathds{R}^{n_l\times 2(n_b-1)}$, $\mathbf{A}_t^{l} \in \mathds{R}^{2 \times 2(n_b-1)}$, $\mathbf{b}_t^\mathbf{v} \in \mathds{R}^{(n_b-1)} $, $\mathbf{b}_t^{\mathbf{i}}  \in \mathds{R}^{n_l}$, and $\mathbf{b}_t^{l} \in \mathds{R}^2$ are parameters for linearizations determined by SCs for time index $t$.
They are determined with the method in \cite{christakou2013efficient_short} by solving a system of linear equations (that admits a single solution, as proven in \cite{paolone2015optimal}) as a function of the grid's admittance matrix and the knowledge of the system state. In the day-ahead phase, the SCs along the whole scheduling horizon are calculated using point predictions of the nodal injections. In the real-time phase, the SCs are updated at each control step using the most recent information on the grid state.
\section{Day-ahead dispatch} \label{sec:Day_ahead}
\subsection{Design requirements of the dispatch plan}
The objective of the day-ahead scheduling is to compute the dispatch plan, mainly the active power trajectory that the targeted distribution network should follow at its GCP during operations. The design requirements of the dispatch plan are: 
\begin{itemize}
    \item stochastic variations from the dispatch plan due to distributed generation and demand should be compensated by the controllable resources while respecting their operational constraints;
    \item the regulation made by the controllable resources does not violate grid constraints;
    \item the power factor at the GCP should be near unity.
\end{itemize}
The dispatch plan is computed with a stochastic optimization framework, where the stochastic injections of distributed generation and demand are modelled through forecast scenarios. Grid constraints are modelled with the linearized grid model discussed in the previous section. Operational constraints of the controllable resources are modelled accounting for the PQ capability of their power converters and state-of-energy constraints. 
\subsection{Computation of the dispatch plan}
Let $r = 1, \dots, R$ be the index of the controllable resources that can participate to the dispatch, $\mathcal{T} = [t_0, t_0+1 \dots, t_{N}]$ the set of time indices of the scheduling horizon delimited by $t_0$ and $t_N$. The set $\Omega$ collects the scenario $\omega$ for stochastic uncontrollable generation and demand.
The active and reactive nodal power injections of controllable and uncontrollable resources for scenario $\omega$ are denoted by  $\mathbf{p}_t^{\omega}, \mathbf{q}_t^{\omega}$ and $\mathbf{p}^{\text{unc},\omega}_t$, $\mathbf{q}^{\text{unc}, \omega}_t$ respectively, where those last are from scenario forecasts. The nodal injections of the controllable resources are the decision variables of the problem and are collected in the vector $x^{\omega}_{r,t} = [{p}^{\omega}_{r,t}, {q}^{\omega}_{r,t}]$.
Let $s_{0,t}^{\omega}$ the complex power at the slack bus for time $t$ and scenario $\omega$. 
Let the complex number $s^{\text{disp}}_t = p^{\text{disp}}_t + j q^{\text{disp}}_t$ the decision variable for the dispatch plan for time $t$, where $p^{\text{disp}}_t$ and $q^{\text{disp}}_t$ refer to the active and reactive power respectively.  

The main idea behind the proposed formulation is to determine a dispatch plan that can be tracked for any of the forecast scenarios. The problem consists in determining the injections of the controllable resources so as to minimize the deviation between the dispatch plan $s^{\text{disp}}$ and slack power for all the scenarios $s_{0}^{\omega}, \omega \in \Omega$.
{
Moreover, the cost function includes a resource-specific term $f^D_r(x_{r,t})$ that reflects the willingness of each controllable resource to provide regulating power (specific cost functions are described in Sec~\ref{sec:Day-ahead_stagee}) and a coefficient $\lambda_r$ to weight them. Both the cost function and the coefficient can be designed by the modeller, for instance, based on a combination of resource’s operating conditions (such as minimize wear and tear, power ramping, power variations etc.) or the monetary cost associated to it's operation. The cost function should be convex in order to keep the convexity of the overall problem formulation. 
The impact of the term $\lambda_r$ is investigated in the Appendix.
}
The problem that we solve is:
\begin{subequations}
\label{eq:day_ahead_form}
\begin{align}
\begin{aligned}
    \hat{s}^{\text{disp}} = \underset{\begin{matrix}
\scriptstyle s^{\text{disp}}\\
\end{matrix}}{\text{arg min}} \sum_{\omega \in \Omega}\sum_{t \in \mathcal{T}} \Big\{(& s_{0,t}^{\omega} - s^{\text{disp}}_t)^2 + \sum_{r=1}^R\lambda_r f^D_r(x_{r,t}^{\omega})\Big\} \label{eq:costdispatch}
\end{aligned}
\end{align}
subject to the power flow at the GCP as a function of the nodal injections and losses
\begin{align}
&  {p}_{0,t}^{\omega}  = \sum_{r=1}^R p_{r,t}^{\omega} + \mathbf{1}^T\mathbf{p}^{\text{unc}, \omega}_t +  p^{l, \omega}_t  ~\forall t \in \mathcal{T}, \omega \in \Omega \label{eq:tracking_error_pdis},\\
&  {q}_{0,t}^{\omega} = \sum_{r=1}^R  q_{r,t}^{\omega} + \mathbf{1}^T\mathbf{q}^{\text{unc}, \omega}_t + q^{l, \omega}_t  ~\forall t \in \mathcal{T}, \omega \in \Omega \label{eq:tracking_error_qdis}, \\
& \begin{bmatrix}
p^{l, \omega}_t~
q^{l, \omega}_t
\end{bmatrix}^T  = 
\mathbf{A}^{l, \omega}_t
\begin{bmatrix}
    \mathbf{p}^{\omega}_t~
    \mathbf{q}^{\omega}_t
\end{bmatrix}^T +
\mathbf{{b}}^{l, \omega}_t ~\forall t \in \mathcal{T}, \omega \in \Omega \label{eq:loss_model_disp},
\end{align}
power factor constraint at the GCP imposed by $\cos(\theta)_{\text{min}}$
\begin{align}
& {|{p}_{0,t}^{\omega}|}/{||{s}_{0,t}^{\omega}||} \geq  {\cos(\theta)_{\text{min}}} & ~\forall t \in \mathcal{T}, \omega \in \Omega \label{eq:Q_cons_disp},
\end{align}
linear voltage and current constraints ($v^{\text{min}},v^{\text{max}}$ are voltage limits, and $\mathbf{i}^{\text{max}}$ lines' ampacities)
\begin{align}
& v^{\text{min}} \leq \mathbf{A}^{\mathbf{v},\omega}_t \begin{bmatrix}
    \mathbf{p}^{\omega}_t~
    \mathbf{q}^{\omega}_t
\end{bmatrix}^T 
+ \mathbf{b}^{\mathbf{v}, \omega}_t \leq v^{\text{max}}  ~\forall t \in \mathcal{T}, \omega \in \Omega \label{eq:volt_const_disp},\\
    & 0 \leq \mathbf{A}^{\mathbf{i}, \omega}_t \begin{bmatrix}
    \mathbf{p}^{\omega}_t~
    \mathbf{q}^{\omega}_t
\end{bmatrix}^T 
+ \mathbf{b}^{\mathbf{i}, \omega}_t \leq \mathbf{i}^{\text{max}}  ~~~\forall t \in \mathcal{T}, \omega \in \Omega \label{eq:current_const_disp},
\end{align}
and constraints for all controllable resources
\begin{align}
\Phi^D_r({x}^{\omega}_{r,t}) & \leq 0 ~~\forall t \in \mathcal{T}, \omega \in \Omega, r = 1, \dots, R \label{eq:local_cons_disp}.
\end{align}
\end{subequations}
Once the problem in \eqref{eq:day_ahead_form} is solved, the dispatch plan is the real part of its solution $\hat{s}^{\text{disp}}$:
\begin{align}
\hat{p}^{\text{disp}} = \Re\left\{\hat{s}^{\text{disp}}\right\}. \label{eq:dispatchplan}
\end{align}

\subsection{Relaxation of the non-convex power factor constraint}
Eq. \eqref{eq:Q_cons_disp} is non-convex and infeasible when the real power at the GCP is zero. As proposed in \cite{stai2018dispatching_short}, we express the active power at the GCP as
\begin{align}
p_{0,t}^{\omega} = p_{0,t}^{+, \omega} - p_{0,t}^{-, \omega} \label{eq:relaxation}
\end{align}
and replace Eq.~\eqref{eq:Q_cons_disp} with the following set of linear constraints:
\begin{align}
&{p_{0,t}^{+,\omega}} + {p_{0,t}^{-,\omega}} \ge q_{0,t}^{\omega}\tan(\pi/2 -\theta_m) \label{eq:pf1}\\
&{p_{0,t}^{+,\omega}} + {p_{0,t}^{-,\omega}} \ge -q_{0,t}^{\omega}\tan(\pi/2 -\theta_m) \label{eq:pf2}\\
&{p_{0,t}^{+,\omega}} \ge 0, {p_{0,t}^{-,\omega}} \ge 0\label{eq:pf4},
\end{align}
for all $t \in \mathcal{T}, \omega \in \Omega$, where $\theta_m$ refers to the angle corresponding to $\cos(\theta)_{\text{min}}$. The two terms of \eqref{eq:relaxation} ($p_{0,t}^{+, \omega},p_{0,t}^{-, \omega}$) should be mutually exclusive. To this end, we augment the cost function \eqref{eq:costdispatch} with the following new term 
\begin{align}
\sum_{\omega \in \Omega}\sum_{t \in \mathcal{T}}\nu\big(({p_{0,t}^{+,\omega}})^2 +({p_{0,t}^{-,\omega}})^2\big)
\end{align}
that promotes $p_{0,t}^{+, \omega},p_{0,t}^{-, \omega}$ being mutually exclusive, where $v\geq0$ weighs the significance of obeying power factor constraints.
\section{Real-time operation}\label{sec:Real-time}
{In the following, we describe the real-time control problem for tracking the day-ahead dispatch plan. Its objective is to determine the set-point for the controllable resources to track the dispatch plan while respecting the grid and resources constraints.
Since the problem requires the knowledge of the state of the grid (i.e., nodal voltages and line currents), of all the components, and power flow at the grid connection point, the problem is initially formulated as a centralized MPC. We then acknowledge that the problem can be solved by consensus among multiple sub-problems as in \cite{6920041, boyd2011distributed_copy}, and we derive a distributed formulation solved by means of the ADMM technique.} 
\paragraph{Centralized MPC} 
\label{sec:centralized_mpc}
During real-time operations, the controllable resources are controlled so to track the dispatch plan at the GCP. The decision variable for the active and reactive nodal injection for resources $r$ at time $t$ is denoted by $x_{r,t} = [{p}_{r,t}, {q}_{r,t}]$ and collected in $\boldsymbol{x}_r = [x_{r,\underline{t}}, \dots, x_{r,t_H}]$ for the length of the optimization horizon delimited by current time interval $\underline{t}$ and the control horizon $t_H$. Let $f_r(x_{r,t})$ denote the actuation cost of {a generic resource $r$ (the specific cost functions are described in Sec~\ref{sec:RT})} and $\hat{p}_t^{\text{disp}}$ the dispatch plan set-point at time $t$ from \eqref{eq:dispatchplan}. The real-time control problem is formulated as MPC. The problem consists in minimizing the actuation costs of the resources subject to their operational constraints, dispatch plan objective, local grid constraints, and power factor limitations over the length of the optimization horizon. The problem is: 
\begin{subequations}\label{eq:centralized1}
\begin{align}
&  ~~~  \underset{\boldsymbol{x}_1, \dots, \boldsymbol{x}_R} {\text{min}}~\sum_{r=1}^R\sum_{t= \underline{t}+1}^{t_H} f_r(x_{r,t})
\end{align}
subject to the dispatch constraint
\begin{align} 
    &  \hat{p}^{\text{disp}}_t = \sum_{r=1}^R p_{r,t} + \mathbf{1}^T\mathbf{p}^{\text{unc}}_t + p^l_t && t = \underline{t}+1, \dots, t_H \label{eq:tracking_error1}
\end{align}
the power factor constraint at the GCP, imposed by $\cos{(\theta)_\text{min}}$
\begin{align}
    & {q}^{\text{gcp}}_t = \sum_{r=1}^R q_{r,t} + \mathbf{1}^T\mathbf{q}^{\text{unc}}_t +q^l_t  && t = \underline{t}+1, \dots, t_H \label{eq:tracking_error_q}\\
    & {|q_t^{\text{gcp}}|} \leq  \frac{|\hat{p}_t^{\text{disp}}|}{\tan(\pi/2-\theta_m)} && t = \underline{t}+1, \dots, t_H \label{eq:Q_cons}
\end{align}
the constraints for all controllable resources
\begin{align}
    & \Phi_r({x}_{r,t}) \leq 0 && r = 1, \dots, R, ~t = \underline{t}+1, \dots, t_H  \label{eq:local_cons}
\end{align}  
and the constraints of the grid
\begin{align}
& \eqref{eq:volt_const_disp}, \eqref{eq:current_const_disp} && t = \underline{t}+1, \dots, t_H \label{eq:grid_dmpc}\\
& \eqref{eq:lin_grid_model_pl}&& t = \underline{t}+1, \dots, t_H.\label{eq:grid_loss}
\end{align} 
\end{subequations}
The formulation in \eqref{eq:centralized1} is convex thanks to the quadratic objective and linear constraints. 
{For convenience in the following formulation, we denote the inequality \eqref{eq:Q_cons}, \eqref{eq:grid_dmpc} and equality \eqref{eq:tracking_error_q}, \eqref{eq:grid_loss} constraints with  $\Psi_{\text{ineq}}(\boldsymbol{x}_1, \dots \boldsymbol{x}_R) \leq 0 $ and $\Psi_{\text{eq}}(\boldsymbol{x}_1, \dots \boldsymbol{x}_R) = 0 $, respectively.}

Solving the problem in \eqref{eq:centralized1} entails knowing the individual resource models and accessing their state during real-time operations. It is, therefore, referred to as centralized. Due to the privacy and security concerns for the resources' owners, the centralized approach may be impractical. For this reason, we resort to a distributed formulation that also assures better scalability with respect to the number of controllable resources.
\paragraph{Distributed MPC with ADMM}
\label{Sec:Distributed_MPC}
As in \cite{gupta2019performance}, a barrier function $g$ is zero cost when the tracking error in \eqref{eq:tracking_error1} is respected and infinity otherwise. We  introduce a set of auxiliary variables $\boldsymbol{z}_r$ that mimic the behaviour of the original variables $\boldsymbol{x}_r$, allowing the centralized problem to be separable in $\boldsymbol{x}_r$. 
Finally, using a sequence of Lagrangian multipliers, denoted by $\boldsymbol{y}_r$, and 
using scaled version of the ADMM sharing problem, the centralized problem \eqref{eq:centralized1} can be solved in following three iterative updates.
\subsubsection{Original variables update} $\boldsymbol{x}_r^{k+1} := $
\begin{subequations}\label{eq:original_updates}
\begin{align}
\underset{\boldsymbol{x}_r} {\text{arg min}}\Big\{\sum_{t = \underline{t}+1}^{t_H}f_r(x_{r,t})  + \frac{\rho}{2}\left|\left|\boldsymbol{x}_{r}- \boldsymbol{z}_{r}^k+ \boldsymbol{u}_r^k\right|\right|_2^2\Big\}
\end{align}
subject to
\begin{align}
    & \Phi_r(x_{r,t}) \leq 0 
    && t = \underline{t}+1, \dots, t_H.
\end{align}
\end{subequations}
\subsubsection{Copied variables update} $[\boldsymbol{z}_{1}^{k+1}, \dots, \boldsymbol{z}_R^{k+1}] := $
\begin{subequations}\label{eq:copied_updates}
\begin{align}
\begin{aligned}
 \underset{\boldsymbol{z}_1 \dots \boldsymbol{z}_R} {\text{arg min}}\Bigg\{ & \sum_{t = \underline{t}+1}^{t_H} \Big\{ g(z_{1,t},\dots,z_{R,t})  \Big\} + \\ 
& +  \frac{\rho}{2}\sum_{r=1}^R\left|\left|\boldsymbol{x}_r^{k+1}- \boldsymbol{z}_r+ \boldsymbol{u}_r^k\right|\right|_2^2\Bigg\}
\end{aligned}
\end{align}
subject to
{
\begin{align}
 & \Psi_{\text{eq}}(\boldsymbol{z}_1,\dots, \boldsymbol{z}_R) = 0 \\
 & \Psi_{\text{ineq}}(\boldsymbol{z}_1,\dots, \boldsymbol{z}_R) \leq 0.
\end{align}}
\end{subequations}
\subsubsection{Dual variable updates}
\begin{figure}
    \centering
    \includegraphics[width=3.3in]{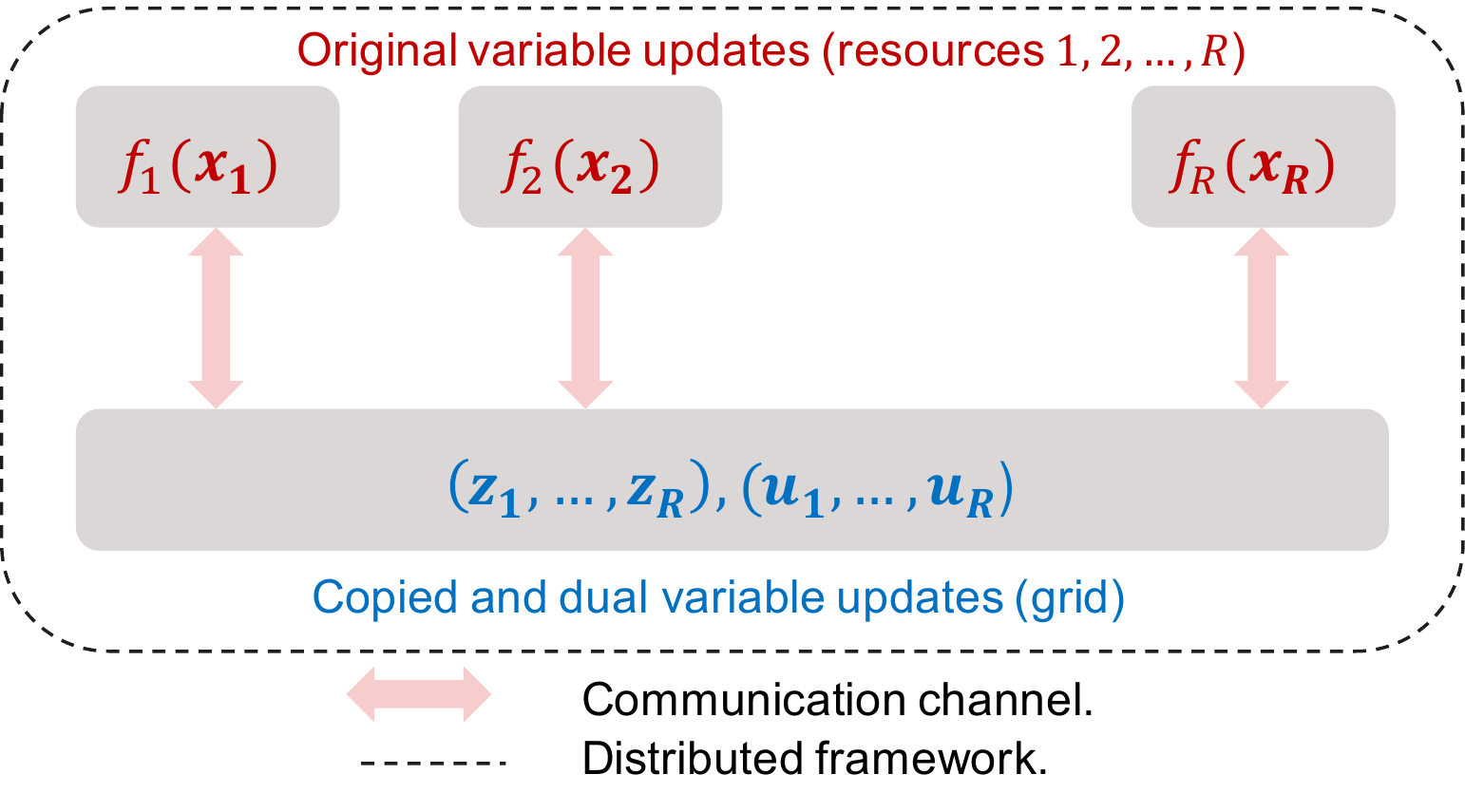}
    \caption{{Distributed computation of the control set-points using the ADMM technique: the resources solve their local problems in parallel and communicates the intermediate set-points to an aggregator that solves the OPF problem associated to the microgrid dispatch. This iterative procedure is followed until convergence.}}
    \label{fig:admm}
\end{figure}

\begin{align}\label{eq:dual_updates}
    \boldsymbol{u}_r^{k+1} = \boldsymbol{u}_r^k + \boldsymbol{x}_r^{k+1} - \boldsymbol{z}_r^{k+1} && r = 1,\dots, R.
\end{align}
Here, $||.||_2$ refers to the euclidean-norm, $k$ is the ADMM iteration index, and $\boldsymbol{u}_r = \boldsymbol{y}_r/\rho$ is a sequence of scaled dual variables, $\rho$ being the standard ADMM penalty parameter. 
The original variable updates (also referred to as resource problems) \eqref{eq:original_updates} are computed in parallel for each resource, $r = 1, \dots, R$, then the copied variable update (grid aggregator problem) \eqref{eq:copied_updates} solves an OPF accounting for the local solutions from each resource, and finally dual variables are updated in \eqref{eq:dual_updates}. Then, the updated solutions of the grid aggregator (copied and dual variables) are sent to the resources for next iteration. Eq. \eqref{eq:original_updates}, \eqref{eq:copied_updates} and \eqref{eq:dual_updates} are solved till convergence.
{The problem in \eqref{eq:original_updates}-\eqref{eq:dual_updates} is distributed because the resource problems \eqref{eq:original_updates} can be solved in parallel and independently without requiring the knowledge of the model of other resources as well as those of the grid. The resource problem requires just the updated solution from the grid (referred to as copied variable) through a communication channel. The diagram of the distributed computation is shown in Fig.~\ref{fig:admm}}.

The convergence criterion is met when the primal residual norm ${r}^{k} = ||\mathbf{X}^k - \mathbf{Z}^k||_2$ and the dual residual norm ${s}^{k} = \rho||\mathbf{Z}^{k+1} - \mathbf{Z}^k||_2$ are both smaller than a feasibility tolerance as in \cite{boyd2011distributed_copy}, where $\mathbf{X} = [\boldsymbol{x}_1; \dots; \boldsymbol{x}_R]$, $\mathbf{Z} = [\boldsymbol{z}_1; \dots; \boldsymbol{z}_R]$. 
For the penalty parameter $\rho$, we follow a self-adaptive scheme \cite{boyd2011distributed_copy, wang2001decomposition_short}: 
\begin{align}\label{eq:adaptive_penalty}
{\rho}^{k+1} :=
\begin{cases} 
    {\tau}_{\text{incr}}{\rho}^{k}  & {r}^{k} > \mu {s}^{k}\\
    {{\rho}^{k}}/{{\tau}_{\text{decr}}}  & {s}^{k} >  \mu {r}^{k}\\
    {\rho}^{k}  & \text{otherwise},
\end{cases}
\end{align}
where $\tau_{\text{incr}}$ and $\tau_{\text{decr}}$ apply an adjustment scheme to
guide the primal and dual residual norms to converge to zero. We fix $\mu$~=~10 and $\tau_{\text{incr}}$~=~2 and $\tau_{\text{decr}}$~=~2 as reported in \cite{wang2001decomposition_short}. 
\section{Experimental Framework} \label{sec:Expt.}
\subsection{Microgrid setup and Distributed Energy Resources}
\begin{figure}[htbp]
    \centering
    \includegraphics[width=3.2in]{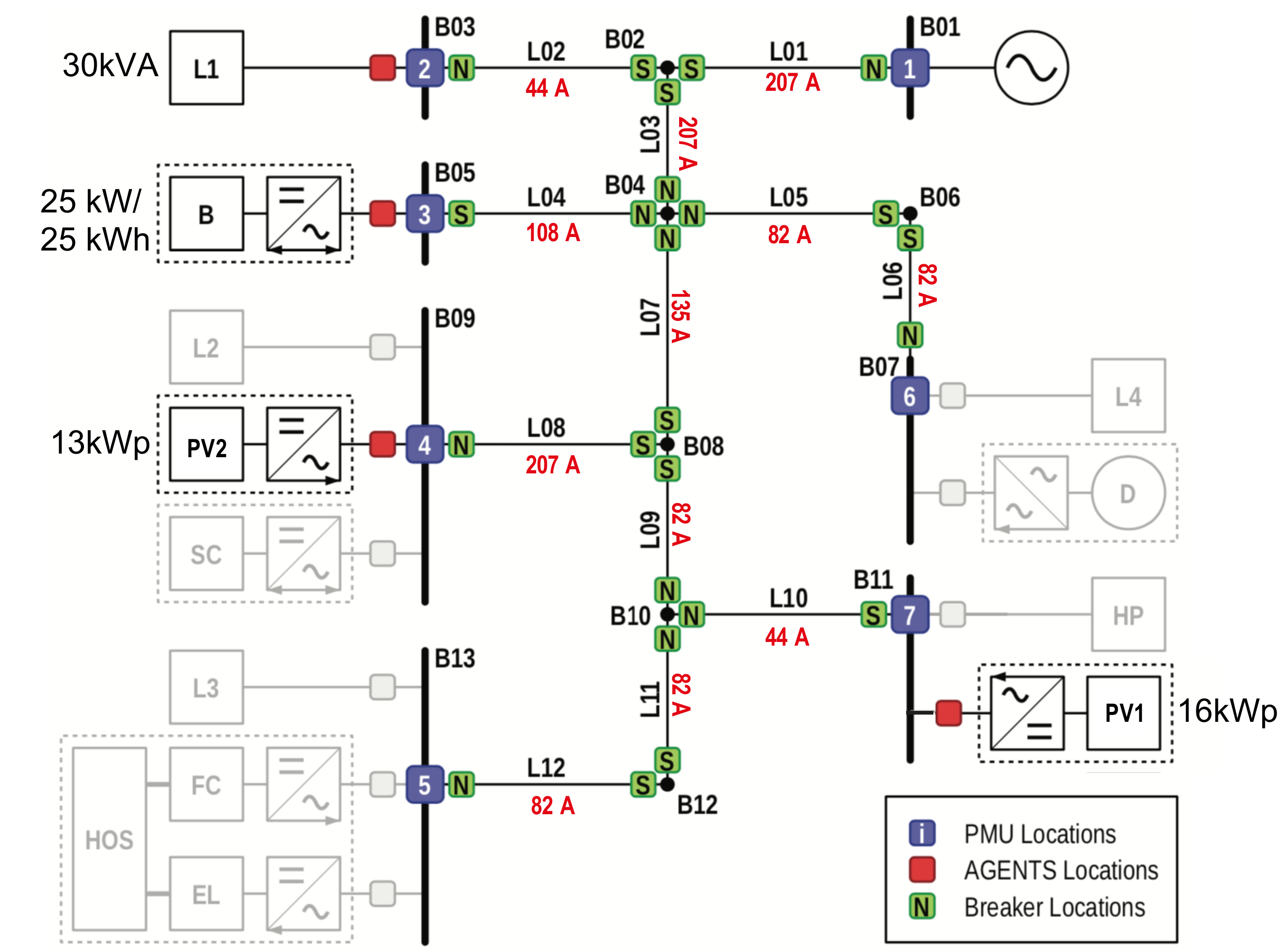}
    \caption{The microgrid setup used for the experimental validation. We consider three controllable resources (a battery B at bus 5, and two curtailable PV plants PV1 and PV2 at bus 11 and 9) and a load emulator (L1 at bus 3).}
    \label{fig:schematic_microgrid}
\end{figure}

\begin{figure}
    \centering
    \includegraphics[width=3in]{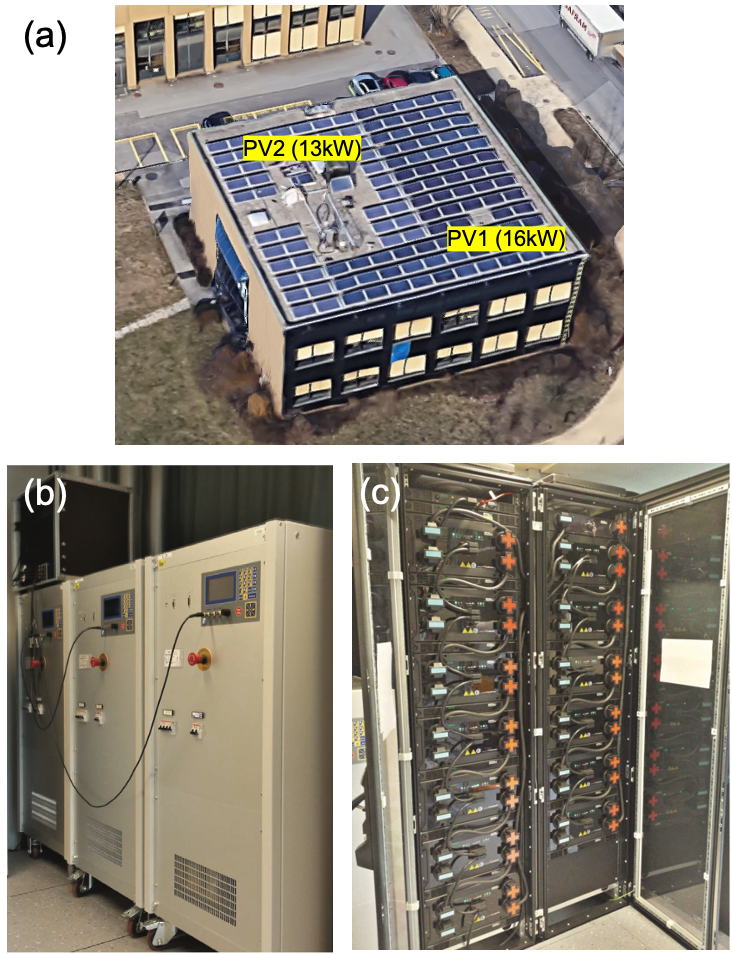}
    \caption{{Experimental setup: (a) Rooftop PV plants PV1 and PV2 (b) Load emulator and (c) Lechlance battery enery storage system. The ratings are reported in Table~\ref{table:Nominal_demand_res}.}}
    \label{fig:resources}
\end{figure}

We validate our control and scheduling algorithms on the real-scale microgrid of the Distributed Electrical Systems Laboratory at EPFL. The setup of the microgrid is inspired by the CIGRE low voltage benchmark microgrid \cite{papathanassiou2005benchmark}. Its setup is shown in Fig.~\ref{fig:schematic_microgrid} that reports the grid topology, the ampacity limits of the cables, locations of the DERs, and the locations of the PMU-based monitoring equipment. The microgrid is operated at 400~V and is connected to a 20~kV medium voltage feeder through a 630~kVA 20/0.4~kV transformer. The microgrid interfaces a number of DERs. For this experimental validation, we consider a load emulator to reproduce stochastic demand, a controllable battery, and two curtailable PV plants, with specifications as reported in Table~\ref{table:Nominal_demand_res}. {They are shown in Fig~\ref{fig:resources}.} The resources are interfaced with the microgrid by power electronics converters. The capability of each resource and associated power converter defines the feasible active and reactive power that can be drawn from each resource.
\\
\begin{table}[h]
\caption{Nominal demands and controllable units}
\renewcommand{\arraystretch}{1.1}
\begin{center}
\begin{tabular}{ |c|c|c|c| } 
\hline
\textbf{node Id} & \textbf{Demand (kVA)} & \textbf{pf} & \textbf{Resource (rating)} \\
  \hline
    B03 & 28 & 0.95 & Load emulator (Zenone)  \\
    \hline
    B05 & -- & -- & BESS (25~kWh/25~kW)\\
    \hline
    B09 & -- & -- & PV2 (13~kWp)\\
    \hline
    B11 & -- & -- & PV1 (16~kWp)\\
   \hline
\end{tabular}\label{table:Nominal_demand_res}
\end{center}
\end{table}
\subsection{Adaptation of the algorithmic framework to the selected DERs}\label{sec:study_case}
We show how to customize the algorithms of sections \ref{sec:Day_ahead} and \ref{sec:Real-time} to fit our experimental setup.
\subsubsection{Day-ahead stage} 
\label{sec:Day-ahead_stagee}
In the day-ahead stage, we consider the battery to be the only controllable resource(\footnote{We do not consider any scheduled curtailment for PV generation. In other words, we let the PV generation work at the maximum power to maximize its capacity factor.}). Its schedule is computed as described in the next paragraph. PV generation and demand are treated as stochastic injections and modelled with forecasts, as discussed later.
\paragraph{Scheduling the operations of the battery} 
battery's active and reactive power schedules $x_{b,t} = [p_{b,t}, q_{b,t}]$ are determined so to respect the power rating of its converter $S^b_{\text{max}}$ and energy capacity $E^b_{\max}$ of the battery. With reference to the formulation in \eqref{eq:day_ahead_form}, the BESS cost function is
\begin{align}
    f^D_b(x_{b,t}) = p_{b,t}^2,
\end{align}
and the constraint set $\Phi^D_b(x_{b,t})$ is
\begin{subequations}
\begin{align}
&\text{SOE}_{t} = \text{SOE}_{t-1} - p_{b,t}T_s  \label{eq:SOE_update_dis}\\
& p_{b,t}^2 + q_{b,t}^2 \leq ({S}^b_{\text{max}})^2   \label{eq:BETT_cap_dis}\\
& aE^b_{\text{max}} \le \text{SOE}_{t} \le (1-a)E^b_{\text{max}} \label{eq:bess_limits_dis}
\end{align}
\end{subequations}
where $\text{SOE}_{t}$ is the BESS state-of-energy, $T_s$ is the sampling time (30~sec), and $0 \leq a < 0.5$ specifies a back-off margin from the SOE limits.
Eq.~\eqref{eq:BETT_cap_dis} constrains the battery's apparent power within its four-quadrant converter capability, which is shown in Fig.~\ref{fig:PQ_profile}(a). We account for BESS power losses by integrating its equivalent series resistance into the network admittance matrix as described in \cite{stai2018dispatching_short}.

%
\paragraph{Forecast scenarios of PV generation and demand}
PV generation is forecasted starting from hourly point predictions of the global horizontal irradiance (GHI). These are from a weather forecast service provider (meteotest.ch) for the location of the experimental microgrid (46.5183$^{\circ}$ N, 6.5652$^{\circ}$ E). As forecasts do not include prediction intervals, we use the following procedure to infer scenarios (to model uncertainty) and increase their time resolution:
\begin{itemize}[leftmargin=*]
\item starting from historical GHI point predictions and measurements from the microgrid pyranometer, we create a training data set consisting of the 1-day-long series of point predictions and associated GHI realizations;
\item given with the GHI point predictions for the target day, we compute its (norm-2) distance from the historical point predictions, take the nearest $s$ (where $s$ is a parameter and is discussed later) and select the associated GHI realizations as potential scenarios of the GHI.
\end{itemize}
The training data set includes two years of measurements from January 2017 at a time resolution of 30~sec. GHI data are scaled by the clear-sky irradiance to remove daily and seasonal components using the PVlib implementation of the Ineichen's clear-sky model \cite{holmgren2018pvlib_short}. PV generation is computed by transposing the GHI data and applying a physical model of PV generation accounting for the air temperature as in \cite{sossan2019solar_short}.

For the demand, we use the same method as in \cite{sossan2016achieving_short}. It consists in selecting $s$ of 1-day long time series of historical measurements of the demand according to calendar information (working day/weekend, day of the week, the period of the year) of the day for which forecasts are to compute.

The selection of $s$ is beyond the scope of the paper and it is defined by the user. In our case, it has been chosen as a trade-off between availability of historical measurements and forecasting performance. In particular, it was verified that including a larger number of scenarios from historical measurements was leading to a marginal improvement of the predicted densities and a negligible impact on the scheduling performance. If parametric forecasting methods are used instead (where the modeller can choose the number of scenarios to generate), $s$ can be designed to achieve a target robustness level as proposed in \cite{calafiore2006scenario_short}.

\subsubsection{Real-time operation stage}
\label{sec:RT}
In real-time operations, all the controllable resources participate in achieving the dispatch objective by leveraging the distributed MPC presented in Sec.~\ref{Sec:Distributed_MPC}. The local problem customized for the specific resources under consideration are discussed in the following. The resolution of the control action is 30~sec and the length of the optimization rolling horizon is 30~minutes\footnote{The choice of time resolution for the control action and the length of MPC optimization horizon is based on the combined time (ADMM computation + communication + implementation) taken by the real-time controller.}.
\paragraph{Battery} the objective is to compute the active and reactive power set-points $x_{b,t} = [p_{b,t}, q_{b,t}]$ while obeying to power rating and energy capacity limits. With reference to the resource problem in \eqref{eq:original_updates} the battery cost function is
\begin{subequations}
\begin{align}
    f_b(x_{b,t}) = {1}
\end{align}
and the constraint set $\Phi_b(x_{b,t})$ is  
\begin{align}
  &  \eqref{eq:SOE_update_dis}, \eqref{eq:BETT_cap_dis},\eqref{eq:bess_limits_dis}.
\end{align}
\end{subequations}
\paragraph{PV power plants}
\label{sec:PV_model_dis}
PV power plants can accept a control signal to curtail generation and implement a reactive power set-point. However, the curtailment action should be kept at a minimum to avoid an excessive impact on the PV capacity factor. The objective of this problem is determining active and reactive power set-points $x_{g,t} = [p_{g,t}, q_{g,t}]$ so as to minimize the curtailment and operate at near-unity power factor while subject to the apparent power limit ${S}^{g}_{\text{max}}$ of the converter. As PV generation is stochastic, the active power injection $p_{g,t}$ is upper-bounded by the theoretical maximum generation potential, that depends primarily on local irradiance conditions. We denote the upper bound of $p_{g,t}$ by $\widehat{p}_{g,t}$. It is derived from short-term point predictions of the irradiance for the horizon 30~sec-30~min by applying the same physical modelling tool-chain described for the day-ahead stage.
Short-term point predictions for the whole horizon are from averaging measurements over the last 2 minutes interval. While doing this, we assume irradiance persistence, that is often regarded to as the reference forecasting model for very-short term look-ahead times \cite{sossan2019solar_short}.
{In this work, we rely on short-term point predictions that are continuously updated by leveraging real-time measurements. In the case of slower refresh times of the control, one could implement prediction intervals to derive robust decisions as in \cite{lara2018robust, valencia2015robust, hosseinzadeh2015robust} to hedge against longer-term uncertainties.
}
With reference to the resource problem in \eqref{eq:original_updates}, the PV cost function is 
\begin{subequations}\label{eq:PV1}
\begin{align}
    f_g(x_{g,t}) = (p_{g,t}-\widehat{p}_{g,t})^2  + q_{g,t}^2
\end{align}
subject to
the constraint set $\Phi_g(x_{g,t})$ 
\begin{align}
   & p_{g,t}^2 + q_{g,t}^2 \leq  ({S}^{g}_{\text{max}})^2 \label{eq:capability_PV} \\
& 0 \leq p_{g,t} \leq  \widehat{p}_{g,t}\label{eq:PVp_limit}
\end{align}
\end{subequations}
Equations \eqref{eq:capability_PV} and \eqref{eq:PVp_limit} are the constraints on active and reactive power injections that account for power converter's capability and PV generation potential.

\begin{figure}[t]
    \centering
    \includegraphics[width = 3.3in]{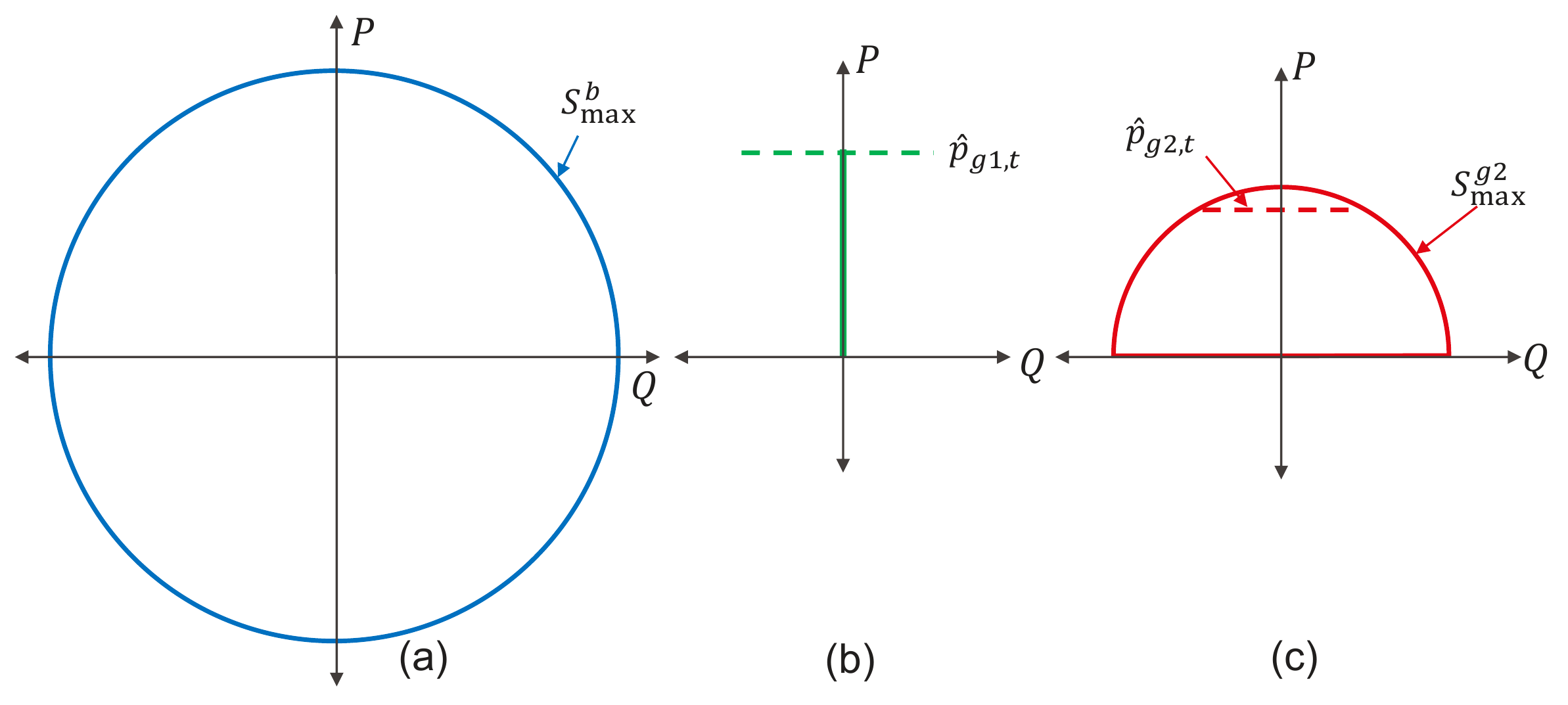}
    \caption{Feasible PQ set for the available DERs: a) battery can be controlled to provide both +/- active (P) and reactive powers (Q), b) PV1 can only provide + P (PV1 converter is not designed to receive external reactive set-points), and c) PV2 can be controlled to provide + P and +/- Q.}
    \label{fig:PQ_profile}.
\end{figure}

In the experiments, we use two curtailable PV units. They differ because they are interfaced to the grid with different power converters. Especially, one of them cannot accept reactive power set-points, whereas the other one can. Their  capability curves are shown in Fig.~\ref{fig:PQ_profile}(b) and \ref{fig:PQ_profile}(c), and are encoded in the constraint \eqref{eq:capability_PV} and \eqref{eq:PVp_limit}.
%
\paragraph{Short-term forecast for the demand}
Given the short look-ahead time, we use a persistent predictor to forecast the demand for the whole optimization horizon.
\subsection{Monitoring and actuation layers}
    \paragraph{Time-deterministic metering infrastructure}
    The real-time control problem requires the knowledge of the grid state at a fast pace (e.g. few sec) to update the linear grid model \eqref{eq:lin_grid_model_v}, \eqref{eq:lin_grid_model_i}, \eqref{eq:lin_grid_model_pl}. 
    Time-synchronised and time-tagged measurements are from Phasor Management Units (PMUs), collected with the setup described in \cite{reyes2018experimental_short, dervivskadic2016architecture_short} that is capable to deliver the measurements at 50 frames per second.
    A discrete Kalman filter-based state estimator processes the measurements \cite{kettner2017sequential} and provides the estimates of the voltage and current phasors of all the nodes and lines with a total latency of less than 80~ms w.r.t. the UTC-GPS time tag of the PMU measurements.
%
    \paragraph{Real-time sensing and processing hardware} Each controllable resource is equipped with a micro-controller (National Instruments CRIO 9068). It is responsible for handling low-level communication tasks such as collecting resource-specific measurements, implementing feasible set-points, and receive set-points from an upper-level controller (i.e., our real-time controller). These functionalities are implemented in LabView.
%
    \paragraph{Communication infrastructure} the microgrid and its resources communicate over a dedicated IPv4 communication network \cite{reyes2018experimental_short}. It provides redundant packets routing using the iPRP redundancy protocol \cite{popovic2016iprp} and secure the messages with the multicast security approach proposed in \cite{tesfay2017experimental_short}. A centralized time series database based on InfluxDB facilitates the exchange of information among the resources and the real-time distributed controllers.
\subsection{Implementation of the algorithms}
\begin{figure}[htbp]
   \centering
    \includegraphics[width = 3.3in]{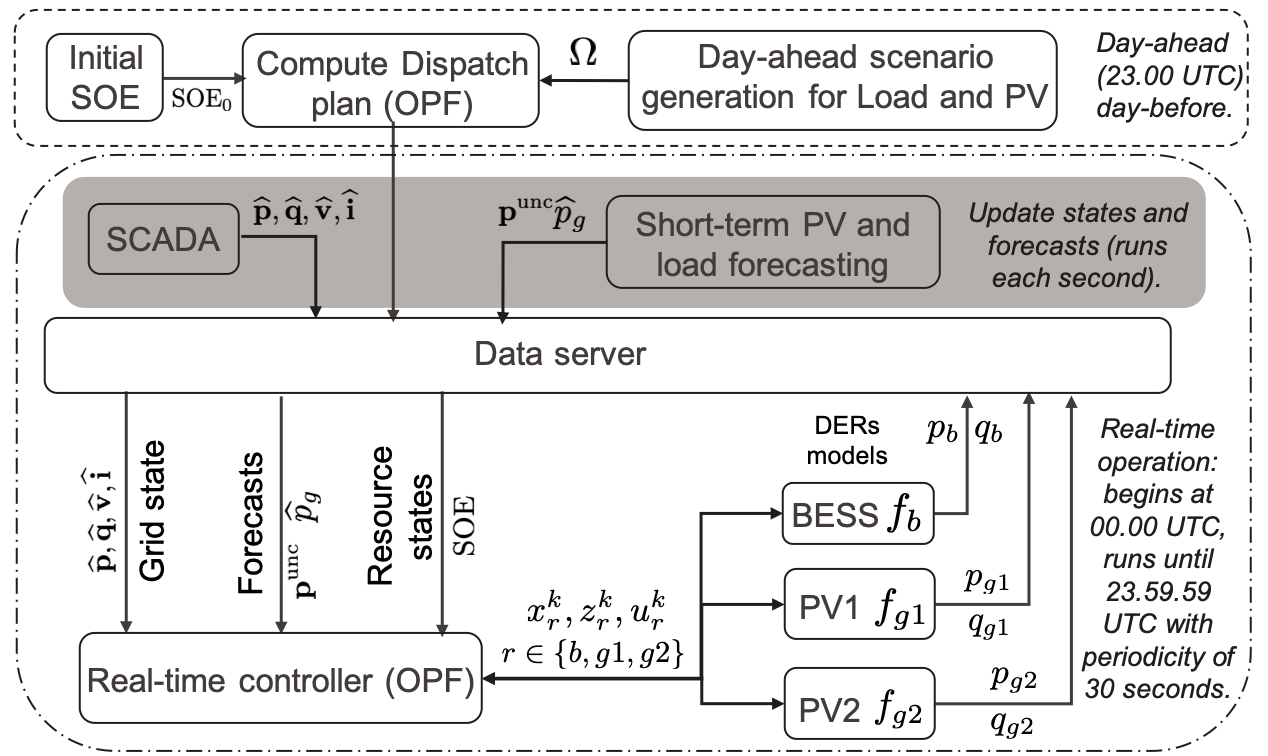}
    \caption{Data flow: Dispatch plan computation starts at 23.00 UTC day-before operation, using the PV and load forecasts, and is stored on the central data-server. Real-time operations start at 00.00 UTC. SCADA and short-term forecasters store their outputs to the data server each second, and ADMM computes power set-points and implements each 30~sec.}
 \label{fig:comm_layout}
\end{figure}

Fig.~\ref{fig:comm_layout} shows the sequence of operations and communication flow among of the day-ahead scheduler and real-time controller. In the former phase (upper dashed rectangle), the dispatch plan is computed and stored in the time series database. In the latter (lower dashed rectangle), a real-time local SCADA, 
the short-term forecasters, and controllable resources save their outputs in the same database (at 1~sec resolution). The real-time controllers access this information to compute the control actions, which are then sent to the controllable resources for actuation through UDP. The set-points are sent continuously to minimize packet losses. The ADMM resource problems are solved in parallel; the intermediate variables are also exchanged through UDP.
\section{Results and discussion}
\label{sec:results_discussion}
We present the experimental results for two days of operations, Day 1 and Day 2, chosen as they feature different PV generation patterns, being characterized by clear-sky and cloudy conditions, respectively.
We focus our analysis on the dispatch plan-tracking performance and the operations of the controllable resources. Grid constraints on nodal voltages and line ampacities are always respected during the experiments and are not shown here because of the limited space.
\subsection{Experimental results}
\subsubsection{Day 1 (4th September 2019)}
\begin{figure}[t]
\centering
\subfloat[Day-ahead net demand scenarios (aggregated demand minus generation).]{\includegraphics[width=3.2in]{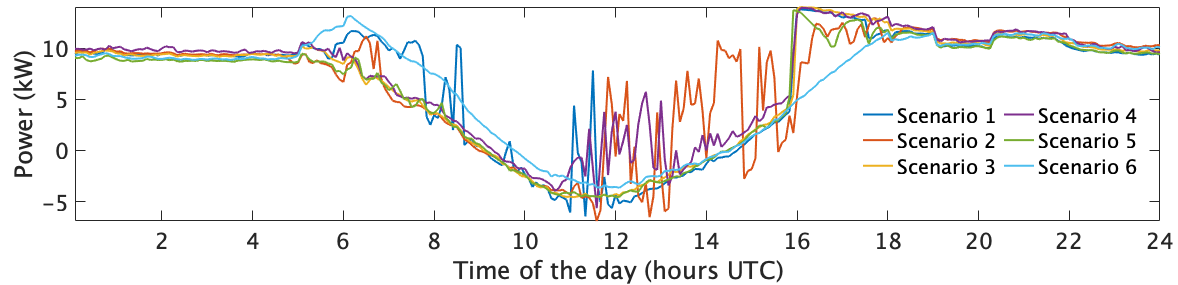}\label{fig:DP_Pros_4sept}}\\
\subfloat[Computed dispatch plan (in red) and scenarios at GCP.]{\includegraphics[width=3.2in]{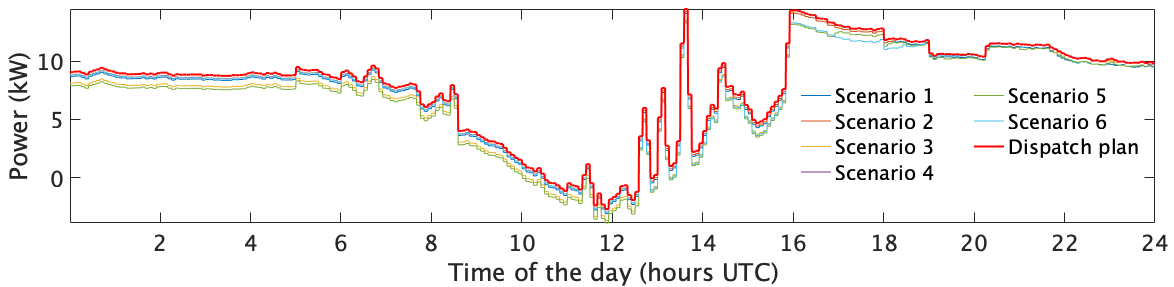}\label{fig:DP_GCP_4sept}} \\
\subfloat[Battery active power injection and SOC for different day-ahead scenarios.]{\includegraphics[width=3.3in]{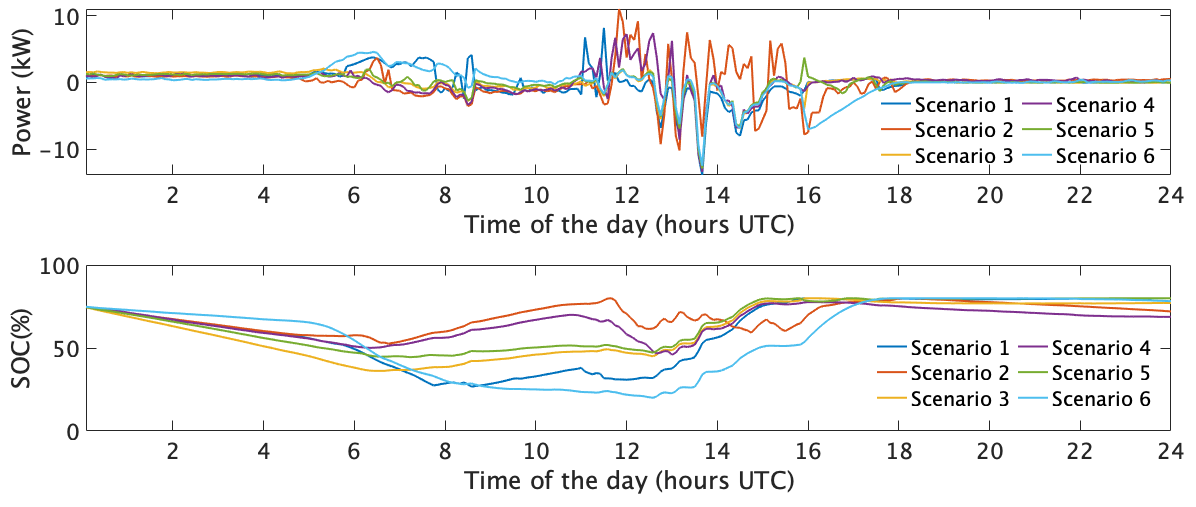}\label{fig:DP_P_SOC_4sept}}
\caption{(a-c) Dispatch plan computation for day 1: 4th September 2019.} \label{fig:Dispatchplan4sept}
\end{figure}

\paragraph{Day-ahead operations}
Fig.~\ref{fig:Dispatchplan4sept} shows the input and output information of the day-ahead dispatch process. The scenarios of the net demand (i.e., aggregated stochastic demand minus generation) at the GCP are shown in Fig.~\ref{fig:DP_Pros_4sept} {which are inputs to the dispatch plan.}
The dispatch plan determined by the algorithm is shown in Fig.~\ref{fig:DP_GCP_4sept} (in red) along with the active power profile scenarios at the GCP (in different colors). The dispatch plan is at a 30~sec resolution. The corresponding battery's power and SOC are shown in Fig.~\ref{fig:DP_P_SOC_4sept}.
As we can see from Fig.~\ref{fig:DP_GCP_4sept}, the dispatch plan appears to be tracked with high fidelity in all the scenarios thanks to the compensation action of the battery. 
Also, to avoid curtailing PV generation and saturating the battery flexibility, the dispatch plan is negative in the central part of the day, denoting that the microgrid exports active power to the upper-level grid. The initial SOC is 0.75, which is the SOC of the battery before the start of the real-time operation. 
\addtolength{\textfloatsep}{-0.2in}
\paragraph{Real-time operations}
The real-time stage starts at 00.00 UTC. The active and reactive power set-points of the controllable resources are computed with the distributed MPC algorithm illustrated in Section~\ref{sec:Real-time}. The dispatch plan is tracked at a 30~sec resolution.

\begin{figure}[t]
\centering
\subfloat[Dispatch plan (in black), measured power at the GCP (in shaded gray) and power at the GCP without distributed MPC (in red).]{\includegraphics[width=3.2in]{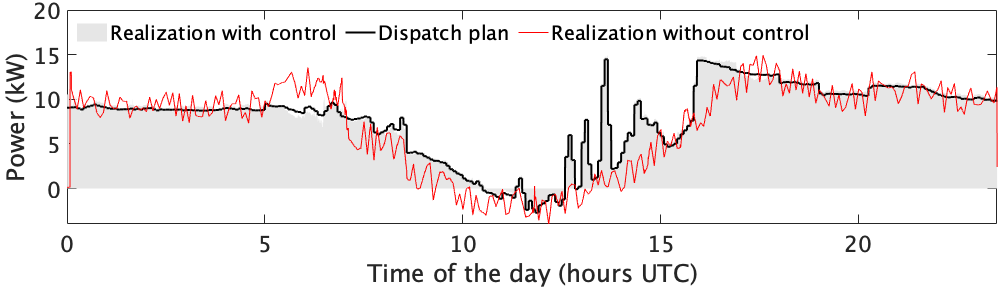}\label{fig:RT_GCP_4sept}}\\
\subfloat[Top: realised battery power injection, bottom: SOC and its limits.]{\includegraphics[width=3.3in]{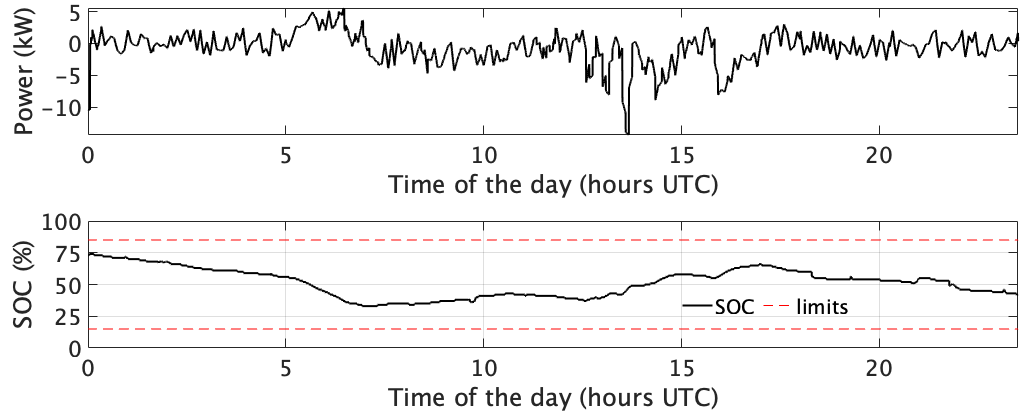}\label{fig:RT_P_SOC_4sept}} \\
\subfloat[Realised generation for PV1 (shaded gray), realised generation for PV2 (shaded green), maximum power for PV1 and PV2 (dashed blue and red).]{\includegraphics[width=3.2in]{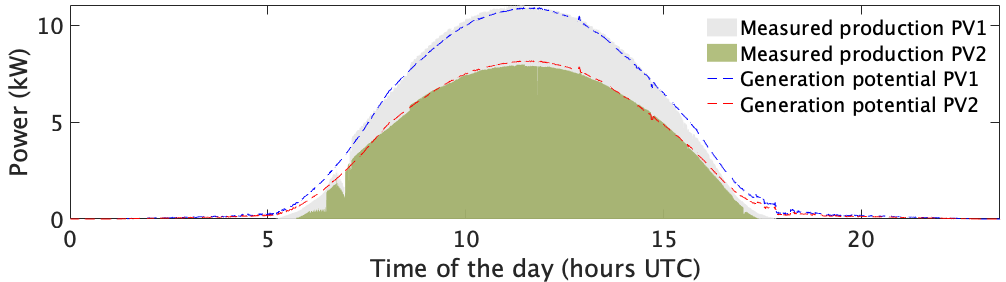}\label{fig:RT_PV_4sept}}
\caption{(a-c) The experimental results for real-time control using the distributed MPC on day~1: 4th September 2019.} \label{fig:RT4sept} 
\end{figure}

The control actions are computed one interval in advance with respect to the actuation time (i.e., 30 sec earlier) and then sent to the resources for being actuated at the designed time interval.
Fig.~\ref{fig:RT_GCP_4sept} shows the power at the GCP with and without the dispatch control action (in shaded gray and solid red), and the dispatch plan (in black). 
As it can be seen, the dispatch plan is tracked with very high fidelity. Fig.~\ref{fig:RT_P_SOC_4sept} shows in the upper panel the battery's active power injection, and SOC evolution in the lower panel.
Fig.~\ref{fig:RT_PV_4sept} shows the measured PV production (in shaded gray and green) and their generation potentials (in dotted blue and red). In this case, there is no curtailment as the battery action alone is sufficient to track the dispatch plan.

To evaluate the dispatch plan-tracking performance, we compute the root mean square error (RMSE), mean, and maximum absolute error (MAE) of the difference between the achieved power at the GCP and the dispatch plan, normalized by the mean of the dispatch plan, with and without control. Results for day~1 are summarised in Table~\ref{tab:dispatch_perf} and show that the control action achieves way better scores than a simple dispatch plan purely based on forecasts. The mean, max and standard deviations (SD) of the time and number of iterations to solve the distributed MPC problem are shown in Table~\ref{tab:compt_stats}. As we can see, the mean and maximum time are well within the 30~sec deadline for the control action actuation.
\subsubsection{Day 2 (10th September 2019)}
\begin{figure}[t]
\centering
\subfloat[Day-ahead net demand scenarios (aggregated demand minus generation).]{\includegraphics[width=3.2in]{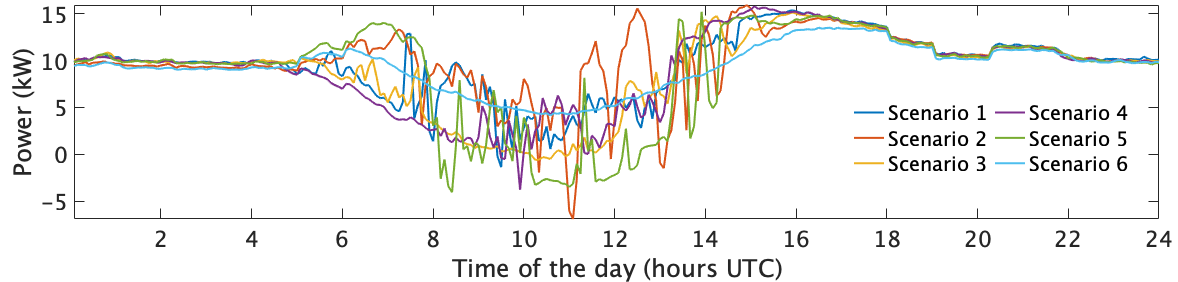}\label{fig:DP_Pros_10sept}}\\
\subfloat[Computed dispatch plan (in red) and scenarios at GCP.]{\includegraphics[width=3.2in]{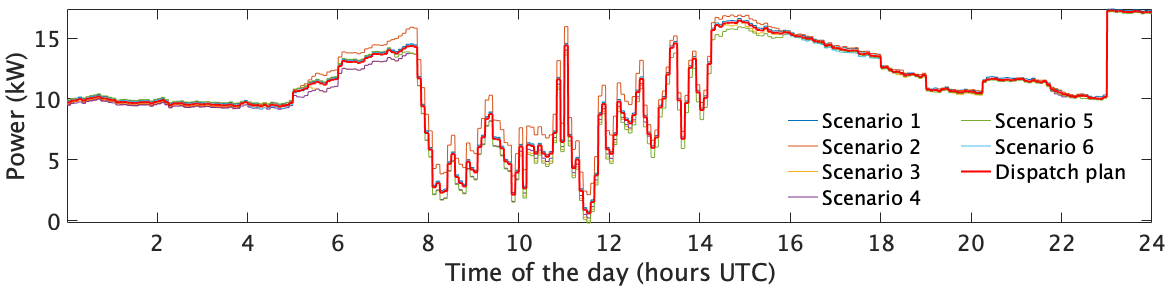}\label{fig:DP_GCP_10Sept}} \\
\subfloat[Battery active power injections and SOC for different day-ahead scenarios.]{\includegraphics[width=3.3in]{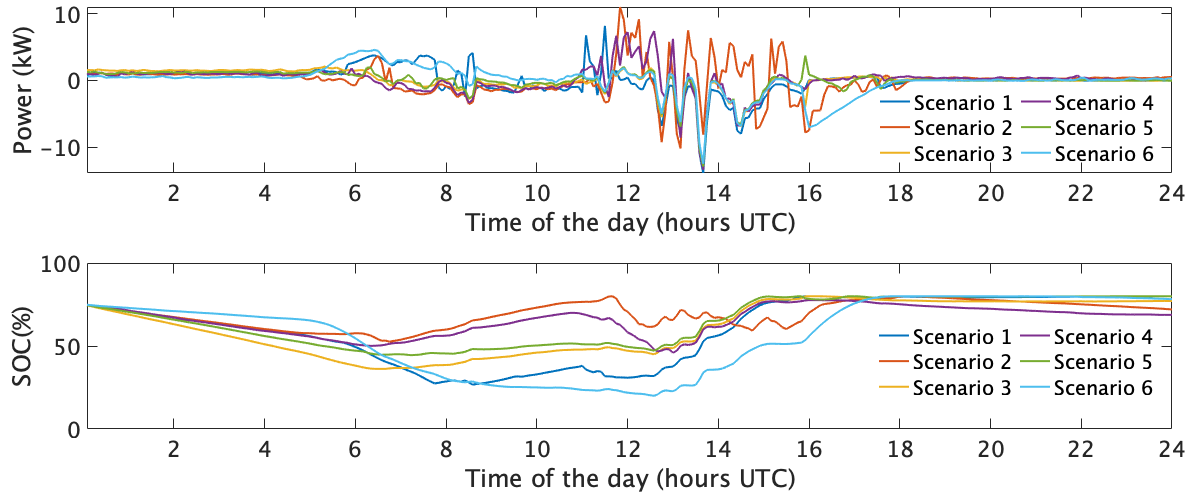}\label{fig:DP_P_SOC_10sept}}
\caption{(a-c) Dispatch plan computation for day~2: 10th September 2019.} \label{fig:Dispatchplan10sept} 
\end{figure}

\paragraph{Day-ahead operations}
The scenario forecasts of the net demand at the GCP are shown in Fig.~\ref{fig:DP_Pros_10sept}.
Compared to day 1 that featured clear-sky conditions, day 2 is partly cloudy and exhibits lower PV generation levels. As a consequence, the dispatch plan, shown in Fig.~\ref{fig:DP_GCP_10Sept}, is positive during all day. {The corresponding battery's power and SOC are shown in Fig.~\ref{fig:DP_P_SOC_10sept}. Again, we can see that the dispatch plan is being tracked with high fidelity in all the scenarios thanks to the compensation action of the battery.}

\begin{figure}[htbp]
\centering
\subfloat[Dispatch plan (in black), measured power at the GCP (shaded gray) and power at the GCP without control (red).]{\includegraphics[width=3.2in]{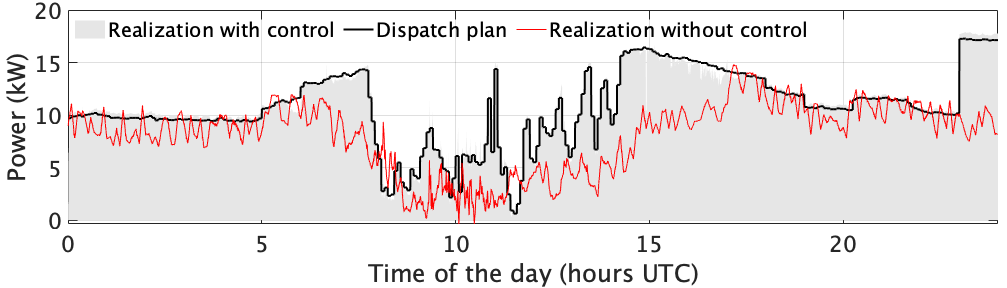}\label{fig:RT_GCP_10sept}}\\
\subfloat[Top: realised battery power injection, bottom: SOC and its limits.]{\includegraphics[width=3.3in]{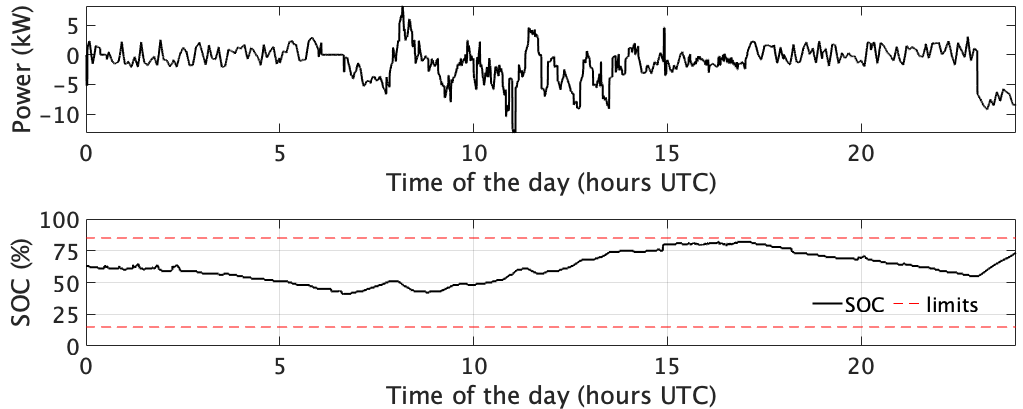}\label{fig:RT_P_SOC_10sept}} \\
\subfloat[Realised generation for PV1 (shaded gray), realised generation for PV2 (shaded green), maximum power for PV1 and PV2 (dashed blue and red).]{\includegraphics[width=3.2in]{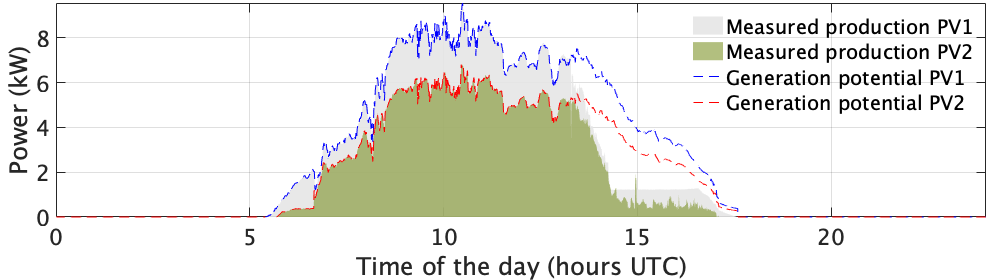}\label{fig:RT_PV_10sept}}
\caption{(a-c) The experimental results for real-time control using the distributed MPC on day~2: 10th September 2019.} \label{fig:RT10sept} 
\end{figure}

\paragraph{Real-time operations} Fig.~\ref{fig:RT_GCP_10sept} shows the power at the GCP with and without the dispatch control action, and the dispatch plan. As visible, the dispatch plan overestimates the net demand in the central part of the day and early afternoon. 

To track the dispatch plan, the controller charges the battery, which, however, approaches a situation of depleted flexibility as it is near the upper SOC limit. As a consequence, the controller curtails both PV power plants starting from 14h, as shown in Fig.~\ref{fig:RT_PV_10sept}. The curtailment action is paramount to follow the dispatch plan at the GCP, which is accurately tracked as visible in Fig.~\ref{fig:RT_GCP_10sept}.

The tracking performance reported in Table~\ref{tab:dispatch_perf} scores slightly worse RMSE than day~1 because of the partly cloudy sky conditions which determine higher PV generation variability. Computation performance in Table~\ref{tab:compt_stats} denotes that the control actions are successfully computed within the 30~sec deadline.
\begin{table}[t]
\centering
\caption{Tracking error statistics with and without dispatch control (in (\%) normalized by mean of the dispatch plan).
}\label{tab:dispatch_perf}
\renewcommand{\arraystretch}{1.1}
\begin{tabular}{| c| l | c |  c | c|  c |  c | c|}
    \hline
    \bf{Scenario} & \multicolumn{3}{|c|}{\textbf{Day 1}} & \multicolumn{3}{|c|}{\textbf{Day 2}}\\
    \hline
    & \bf{RMSE} & \bf{Mean} &\bf{MAE}& \bf{RMSE} & \bf{Mean} &\bf{MAE}  \\
   \hline
    No dispatch & 35.7  & 12.0 &  182.2 & 37.8  & 25.1 & 123.5\\
   \hline
    Dispatch & 4.2 & -1.77 & 23.8 & 4.3 & -0.30 & 21.3 \\
    \hline
\end{tabular}
\end{table}
\begin{table}[t]
\centering
\caption{Computation performance for real-time experiments}\label{tab:compt_stats}
\renewcommand{\arraystretch}{1.1}
\begin{tabular}{| l | c |  c | c | c |  c | c |}
    \hline
    \bf{Day} & \multicolumn{3}{|c|}{\textbf{Time (sec)}} & \multicolumn{3}{|c|}{\textbf{ADMM iterations}}\\
    \hline
     & \bf{Mean} & \bf{SD} &\bf{Max} & \bf{Mean} & \bf{SD} &\bf{Max} \\
   \hline
   1  & 8.10  & 5.34 & 18.56 & 9.34  & 6.60 & 19\\
    \hline
   2 & 5.90  & 4.13 & 13.70 & 7.03 & 5.29 & 17\\
     \hline
\end{tabular}
\end{table}
{
\subsection{Further analysis} This section is devoted to the following analyses: i) we perform a sensitivity analysis on the performance of the distributed approach by increasing the number of controllable elements and ii) we compare the performance (i.e., optimality and computation time) of the centralized vs the distributed formulations.  The analysis is performed by considering the same conditions as day 1. 
\subsubsection{Analysis of the algorithms performance with respect to the number of controllable units} 
This analysis is carried out by dedicated simulations with different number of distributed BESSs. These resources are considered to have identical power rating and total energy capacity equal to the one of the BESS in the experimental validation. The largest number of controllable BESS is equal to 4 units since it appears a reasonable estimate of the possible largest number of BESSs that could be installed in a low-voltage distribution network as the one that we have considered. Indeed, a larger number of BESSs would result in excessively small BESS power ratings (compared to the nominal power of the nodes) and would multiply grid connection costs. The additional BESSs are placed at nodes 5, 6, 7, 8 respectively. Table~\ref{tab:sense_with_bess} reports the corresponding computational performance. Fig.~\ref{fig:time_w_BESS} shows the boxplots of the computation time taken by each resource and the grid for the distributed scheme (the figure refers to a daily time horizon). The total computation time is given by the maximum among the resources plus the grid time. As it can been seen, the average computation time for the BESS increases with the number of units. However, it does not impact the total computation time significantly as the units solve their own problem in parallel. Therefore, increasing the number of controllable BESSs does not influence the solvability of the problem given the real-life solution time constraints.
\begin{table}[t]
\centering
\caption{{Computation time with respect to increasing number of controllable units.}} \label{tab:sense_with_bess}
\renewcommand{\arraystretch}{1.1}
{
\begin{tabular}{| c| c | c | }
    \hline
    \bf{\# BESS units}  & \multicolumn{2}{|c|}{\bf{Total time (sec)}} \\
   \hline
    & \bf{Mean} &\bf{Max} \\
     \hline
   1  & 5.6 & 11.8  \\
   \hline
   2  & 5.6 & 21.3  \\
   \hline
   4   & 4.7 & 12.6  \\
    \hline
\end{tabular}}
\end{table}
\begin{figure}[h]
    \centering
    \includegraphics[width = 3.5in]{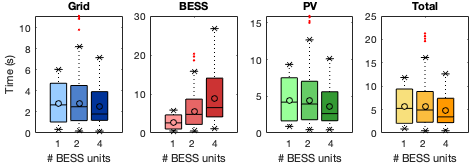}
    \caption{Computation time with number of BESS units for Day 2 (Simulation).}
    \label{fig:time_w_BESS}
\end{figure}
%
\subsubsection{Performance comparison of the centralised vs the decentralised algorithms}
Table~\ref{tab:centVSadmm} shows the results of the comparison in terms of dispatch tracking (measured by RMSE, mean and MAE in $\%$ of the mean of the dispatch plan) and computation time performance of the two proposed algorithms. As visible from Table~\ref{tab:centVSadmm}, the dispatching performance in both cases is very similar. The computation time of the centralized algorithm is shorter than the distributed one. This is to be expected because the distributed optimization formulation requires multiple iterations of the optimization problems to converge to a solution, whereas the centralized algorithm solves a single optimization problem. However, the latter needs to know the complete models of PV and BESS resources, which might not be available in real-life especially when resources belong to different owners. Another advantage of the distributed algorithm is that it is solved by several computers usually characterised by low computing power, whereas the centralized algorithm does require a single computer with significantly larger computing power.
\begin{table}[t]
\centering
\caption{{Performance comparison of the centralized vs distributed algorithms.}} \label{tab:centVSadmm}
\renewcommand{\arraystretch}{1.1}
    {
\begin{tabular}{| l | c | c |c| c|c|}
    \hline
    \bf{Method} & \multicolumn{3}{|c|}{\textbf{Dispatch error}} & \multicolumn{2}{|c|}{\textbf{Time (sec})}\\
    \hline
     & \bf{RMSE} & \bf{Mean} & \bf{Max} &  \bf{Mean} & \bf{Max}  \\
     \hline
   Centralized  & 1.6  & -0.5 & 8.9 & 2.0 & 3.6\\
   \hline
   Distributed  & 3.18  & -2.1 &  10.75 & 5.6 & 11.8\\
     \hline
\end{tabular}}
\end{table}
}
\section{Conclusions}\label{sec:conclusion}
We proposed and experimentally validated a scheduling and control framework for DERs that achieves to track a dispatch plan at the GCP of a distribution network that interconnects heterogeneous resources while respecting constraints on nodal voltages and lines ampacities of the local grid.
In the scheduling phase, we determine an aggregated dispatch plan at the GCP by accounting for forecasts of stochastic generation and demand, the state of the controllable resources, and constraints of the grid. 
During real-time operations, a distributed MPC adjusts the power injections of the controllable DERs to track the dispatch plan subject to grid's and DERs' operational constraints. We leverage a distributed formulation for improved scalability and privacy-preserving properties. To achieve a tractable formulation of the control problem, we used a linearized grid model based on sensitivity coefficients, computed considering point predictions in the scheduling phase, and updated by using the most recent grid state during real-time operations.

The framework is experimentally validated in a real-scale microgrid hosting heterogeneous controllable resources and monitored with PMUs. The dispatch plan, which is at a 30~sec resolution, is computed the day before operations for the next calendar day. The real-time control set-points are implemented every 30~sec for all day. Experimental results, carried out on two distinct days characterized by different irradiance and PV generation patterns, showed that the proposed framework achieves a reliable and accurate dispatch on a 30~sec basis, with RMS and mean tracking errors smaller than 5\% and 2\%, respectively, while respecting all grid constraints.
{
\appendix
\subsection{Sensitivity of dispatch plan quality with weighing coefficient}
\label{sec:sense_w_weights}
We vary the weighing coefficient $\lambda_r$ over a range of values and recompute the dispatch plan. To quantify the reliability of dispatch plan, we use the
mean of RMSE (mRMSE) between each prosumption scenario and the obtained dispatch plan. mRMSE is normalized and expressed in $\%$ of the mean of the dispatch plan. As an example, Table~\ref{tab:weight_coeff} lists the mRMSE for different  $\lambda_r$ for the dispatch computation on Day~2. 
It can be seen that the variation in the mRMSE  for different values of $\lambda_r$, expressed as percentage of the average dispatch plan, is small (less than 4~$\%$ for $\lambda_r$ variation from 0.5e-5 to 0.5) and thus the performance of the proposed problem formulation appears invariant with respect to the value of $\lambda_r$.}
%

\begin{table}[h]
    \centering
    \caption{{Sensitivity of dispatch plan reliability}}
    {
    \begin{tabular}{|c|c|c|c|c|c|}
        \hline
         ${\lambda_r}$ & 0.5e-5 & 0.5-4 & 0.5e-2 & 0.5e-1 & 0.5\\
         \hline
         \bf{mRMSE} & 2.7 & 0.7 & 2.0 & 2.9 & 3.8 \\
         \hline
    \end{tabular}
    }
    \label{tab:weight_coeff}
\end{table}

\bibliographystyle{IEEEtran}
\bibliography{biblio.bib}

\begin{thebibliography}{10}
\providecommand{\url}[1]{#1}
\csname url@samestyle\endcsname
\providecommand{\newblock}{\relax}
\providecommand{\bibinfo}[2]{#2}
\providecommand{\BIBentrySTDinterwordspacing}{\spaceskip=0pt\relax}
\providecommand{\BIBentryALTinterwordstretchfactor}{4}
\providecommand{\BIBentryALTinterwordspacing}{\spaceskip=\fontdimen2\font plus
\BIBentryALTinterwordstretchfactor\fontdimen3\font minus
  \fontdimen4\font\relax}
\providecommand{\BIBforeignlanguage}[2]{{%
\expandafter\ifx\csname l@#1\endcsname\relax
\typeout{** WARNING: IEEEtran.bst: No hyphenation pattern has been}%
\typeout{** loaded for the language `#1'. Using the pattern for}%
\typeout{** the default language instead.}%
\else
\language=\csname l@#1\endcsname
\fi
#2}}
\providecommand{\BIBdecl}{\relax}
\BIBdecl

\bibitem{6417004_short}
R.~{Palma-Behnke} \emph{et~al.}, ``A microgrid energy management system based
  on the rolling horizon strategy,'' \emph{IEEE Trans. Smart Grid}, 2013.

\bibitem{6913566_short}
M.~Abu~Abdullah \emph{et~al.}, ``An effective power dispatch control strategy
  to improve generation schedulability and supply reliability of a wind farm
  using a battery energy storage system,'' \emph{IEEE Trans. Sust. Energy},
  2015.

\bibitem{he2011multiple}
M.~He \emph{et~al.}, ``Multiple timescale dispatch and scheduling for
  stochastic reliability in smart grids with wind generation integration,'' in
  \emph{2011 Proceedings IEEE INFOCOM}.\hskip 1em plus 0.5em minus 0.4em\relax
  IEEE, 2011, pp. 461--465.

\bibitem{sossan2016achieving_short}
F.~Sossan \emph{et~al.}, ``Achieving the dispatchability of distribution
  feeders through prosumers data driven forecasting and model predictive
  control of electrochemical storage,'' \emph{IEEE Trans. Sust. Energy}, 2016.

\bibitem{lorca2014adaptive}
A.~Lorca \emph{et~al.}, ``Adaptive robust optimization with dynamic uncertainty
  sets for multi-period economic dispatch under significant wind,'' \emph{IEEE
  Trans. Power Sys.}, vol.~30, no.~4, pp. 1702--1713, 2014.

\bibitem{Christakou_voltage_short}
K.~Christakou \emph{et~al.}, ``Gecn: Primary voltage control for active
  distribution networks via real-time demand-response,'' \emph{IEEE Trans.
  Smart Grid}, 2013.

\bibitem{Bernstein2015_short}
A.~Bernstein \emph{et~al.}, ``A composable method for real-time control of
  active distribution networks with explicit power setpoints. part i:
  Framework,'' \emph{Electric Power Systems Research}, 2015.

\bibitem{gupta2018admm_short}
R.~Gupta \emph{et~al.}, ``An admm-based coordination and control strategy for
  pv and storage to dispatch stochastic prosumers: Theory and experimental
  validation,'' in \emph{2018 Power Syst. Comput. Conf.}, June 2018.

\bibitem{namor2018control_short}
E.~Namor \emph{et~al.}, ``Control of battery storage systems for the
  simultaneous provision of multiple services,'' \emph{IEEE Trans. Smart Grid},
  2018.

\bibitem{espin2020energy_short}
D.~Esp{\'\i}n-Sarzosa \emph{et~al.}, ``Energy management systems for
  microgrids: Main existing trends in centralized control architectures,''
  \emph{Energies}, 2020.

\bibitem{christakou2013efficient_short}
K.~Christakou \emph{et~al.}, ``Efficient computation of sensitivity
  coefficients of node voltages and line currents in unbalanced radial
  electrical distribution networks,'' \emph{IEEE Trans. Smart Grid}, 2013.

\bibitem{gupta2019performance}
R.~Gupta, F.~Sossan, and M.~Paolone, ``{Performance Assessment of Linearized
  OPF-based Distributed Real-time Predictive Control},'' in \emph{{IEEE
  PowerTech 2019}}, Milan, Italy, Jun. 2019.

\bibitem{papathanassiou2005benchmark}
S.~Papathanassiou \emph{et~al.}, ``A benchmark low voltage microgrid network,''
  in \emph{Proc. CIGRE sympos.: power syst. dispersed gen.}\hskip 1em plus
  0.5em minus 0.4em\relax CIGRE, 2005.

\bibitem{perez2012predictive_short}
E.~Perez \emph{et~al.}, ``Predictive power control for pv plants with energy
  storage,'' \emph{IEEE Trans. Sust. Energy}, vol.~4, no.~2, pp. 482--490,
  2012.

\bibitem{halamay2014improving_short}
D.~Halamay \emph{et~al.}, ``Improving wind farm dispatchability using model
  predictive control for optimal operation of grid-scale energy storage,''
  \emph{Energies}, vol.~7, no.~9, pp. 5847--5862, 2014.

\bibitem{teleke2009control_short}
S.~Teleke \emph{et~al.}, ``Control strategies for battery energy storage for
  wind farm dispatching,'' \emph{IEEE Trans. Energy Conv.}, vol.~24, 2009.

\bibitem{zheng2017distributed_short}
Y.~Zheng \emph{et~al.}, ``Distributed model predictive control for on-connected
  microgrid power management,'' \emph{IEEE Trans. Control Syst. Tech.}, 2017.

\bibitem{du2017distributed_short}
Y.~Du \emph{et~al.}, ``Distributed mpc for coordinated energy efficiency
  utilization in microgrid systems,'' \emph{IEEE Trans. Smart Grid}, 2017.

\bibitem{paolone2015optimal}
M.~Paolone, J.-Y. Le~Boudec, K.~Christakou, and D.-C. Tomozei, ``Optimal
  voltage control processes in active distribution networks,'' The Institution
  of Engineering and Technology-IET, Tech. Rep., 2015.

\bibitem{stai2018dispatching_short}
E.~Stai \emph{et~al.}, ``Dispatching stochastic heterogeneous resources
  accounting for grid and battery losses,'' \emph{IEEE Trans. Smart Grid},
  2018.

\bibitem{6920041}
E.~Dall’Anese, S.~V. Dhople, B.~B. Johnson, and G.~B. Giannakis,
  ``Decentralized optimal dispatch of photovoltaic inverters in residential
  distribution systems,'' \emph{IEEE Transactions on Energy Conversion},
  vol.~29, no.~4, pp. 957--967, Dec 2014.

\bibitem{boyd2011distributed_copy}
S.~Boyd \emph{et~al.}, ``Distributed optimization and statistical learning via
  the alternating direction method of multipliers,'' \emph{Foundations and
  Trends in Machine Learning}, vol.~3, 2011.

\bibitem{wang2001decomposition_short}
S.~Wang and L.~Liao, ``Decomposition method with a variable parameter for a
  class of monotone variational inequality problems,'' \emph{Journal of
  optimization theory and applications}, vol. 109, no.~2, pp. 415--429, 2001.

\bibitem{holmgren2018pvlib_short}
W.~Holmgren, C.~Hansen, and M.~Mikofski, ``pvlib python: A python package for
  modeling solar energy systems,'' \emph{J. Open Source Soft.}, 2018.

\bibitem{sossan2019solar_short}
F.~Sossan \emph{et~al.}, ``Solar irradiance estimations for modeling the
  variability of photovoltaic generation and assessing violations of grid
  constraints: A comparison between satellite and pyranometers measurements
  with load flow simulations,'' \emph{J. Renewable Sustainable Energy}, 2019.

\bibitem{calafiore2006scenario_short}
G.~C. Calafiore and M.~C. Campi, ``The scenario approach to robust control
  design,'' \emph{IEEE Trans. autom. control}, 2006.

\bibitem{lara2018robust}
J.~D. Lara, D.~E. Olivares, and C.~A. Ca{\~n}izares, ``Robust energy management
  of isolated microgrids,'' \emph{IEEE Systems Journal}, vol.~13, no.~1, pp.
  680--691, 2018.

\bibitem{valencia2015robust}
F.~Valencia, J.~Collado, D.~S{\'a}ez, and L.~G. Mar{\'\i}n, ``Robust energy
  management system for a microgrid based on a fuzzy prediction interval
  model,'' \emph{IEEE Transactions on Smart Grid}, vol.~7, no.~3, pp.
  1486--1494, 2015.

\bibitem{hosseinzadeh2015robust}
M.~Hosseinzadeh and F.~R. Salmasi, ``Robust optimal power management system for
  a hybrid ac/dc micro-grid,'' \emph{IEEE Transactions on Sustainable Energy},
  vol.~6, no.~3, pp. 675--687, 2015.

\bibitem{reyes2018experimental_short}
L.~Reyes-Chamorro \emph{et~al.}, ``Experimental validation of an explicit
  power-flow primary control in microgrids,'' \emph{IEEE Trans. Ind.
  Informat.}, 2018.

\bibitem{dervivskadic2016architecture_short}
A.~Dervi{\v{s}}kadi{\'c} \emph{et~al.}, ``Architecture and experimental
  validation of a low-latency phasor data concentrator,'' \emph{IEEE Trans.
  Smart Grid}, 2016.

\bibitem{kettner2017sequential}
A.~M. Kettner \emph{et~al.}, ``Sequential discrete kalman filter for real-time
  state estimation in power distribution systems: Theory and implementation,''
  \emph{IEEE Trans. Inst. Meas.}, vol.~66, no.~9, pp. 2358--2370, 2017.

\bibitem{popovic2016iprp}
M.~Popovic \emph{et~al.}, ``iprp—the parallel redundancy protocol for ip
  networks: Protocol design and operation,'' \emph{IEEE Trans. Ind. Informat.},
  vol.~12, no.~5, pp. 1842--1854, 2016.

\bibitem{tesfay2017experimental_short}
T.~T. Tesfay \emph{et~al.}, ``Experimental comparison of multicast
  authentication for wide area monitoring systems,'' \emph{IEEE Trans. Smart
  Grid}, 2017.

\end{thebibliography}


\begin{IEEEbiography}[{\includegraphics[width=1in,height=1.25in,keepaspectratio]{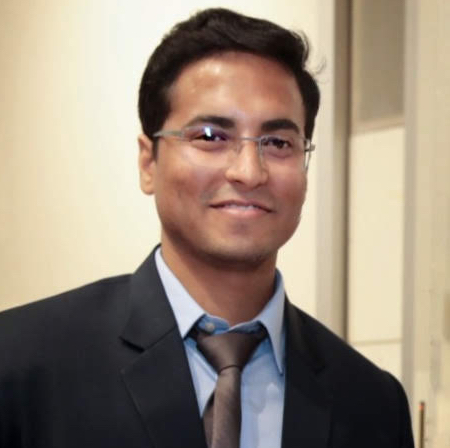}}]{Rahul Gupta} completed his B.Tech in electrical engineering at NIT Rourkela, India in 2014. From 2015 to 2016, he worked as research assistant on micro-energy harvesting at NUS Singapore. In 2018, he received his M.Sc degree in electrical engineering with orientation in smart grids technology at EPFL Lausanne, Switzerland. He received Zanelli technologie et développement durable prize 2018 for his master project in the field of sustainable development. Currently, he is pursuing his Ph.D. degree at the Distributed Electrical Systems Laboratory, EPFL. His research interests include model predictive control, distributed optimization and data-driven control of the active distribution networks in the presence of uncertainties.
\end{IEEEbiography}
\begin{IEEEbiography}[{\includegraphics[width=1in,height=1.25in,clip,keepaspectratio]{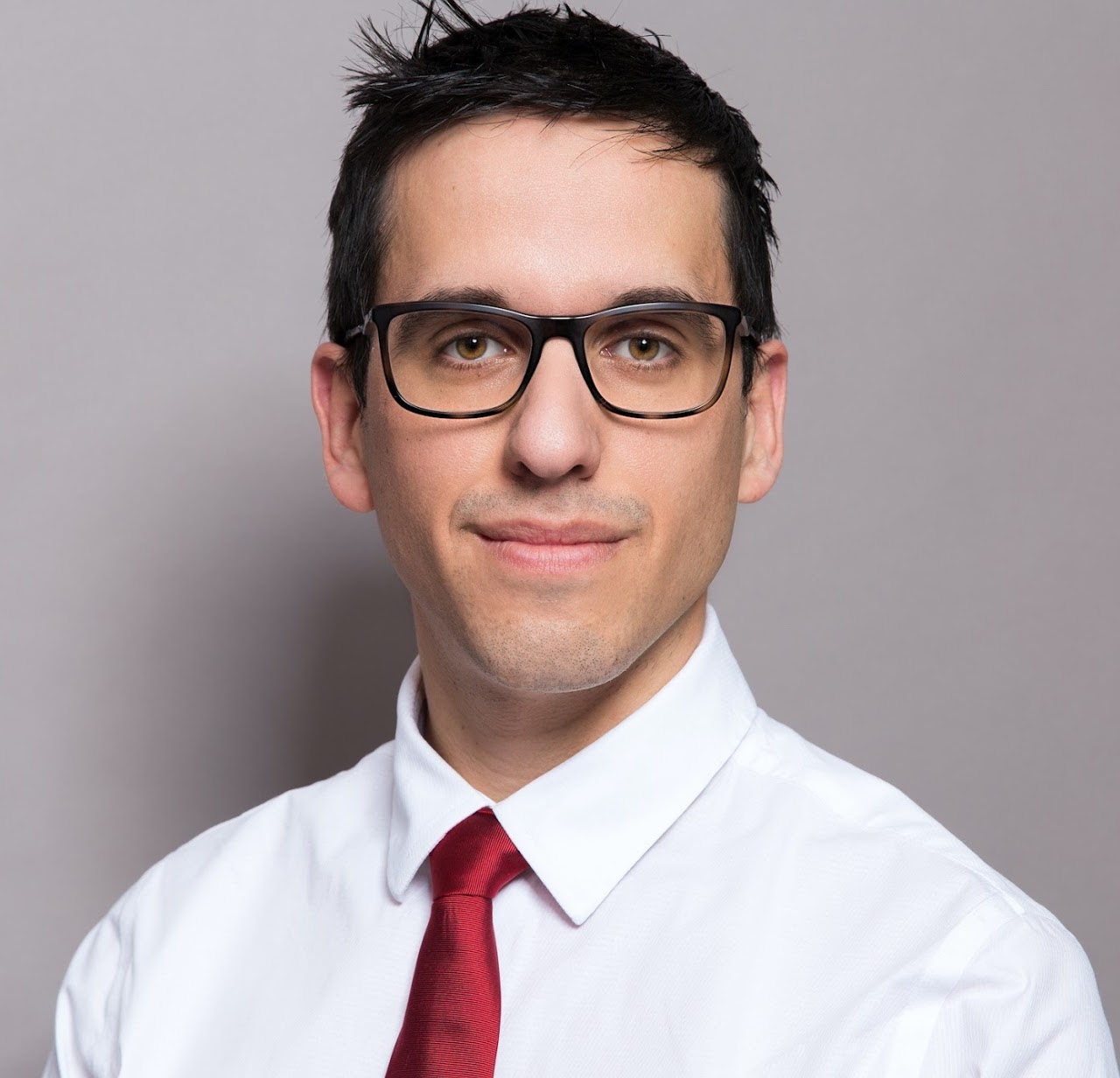}}]{Fabrizio Sossan} completed his studies in computer engineering at the University of Genova in 2010, and he obtained his Ph.D. in Electrical Engineering from the Technical University of Denmark (DTU) in 2014. From 2014 to 2017, he was a postdoctoral fellow at EPFL Lausanne in Switzerland, in 2018 a guest scientist at NREL in the USA and a scientist at ETH in Switzerland. From 2019, he is a senior researcher and lecturer at MINES ParisTech in France. His research interests are modeling and operation of distributed energy resources and energy storage, including forecasting, for renewable power grids. He is an associate editor of Elsevier Sustainable Energy Grids and Networks.
\end{IEEEbiography}
\begin{IEEEbiography}[{\includegraphics[width=1in,height=1.25in,clip,keepaspectratio]{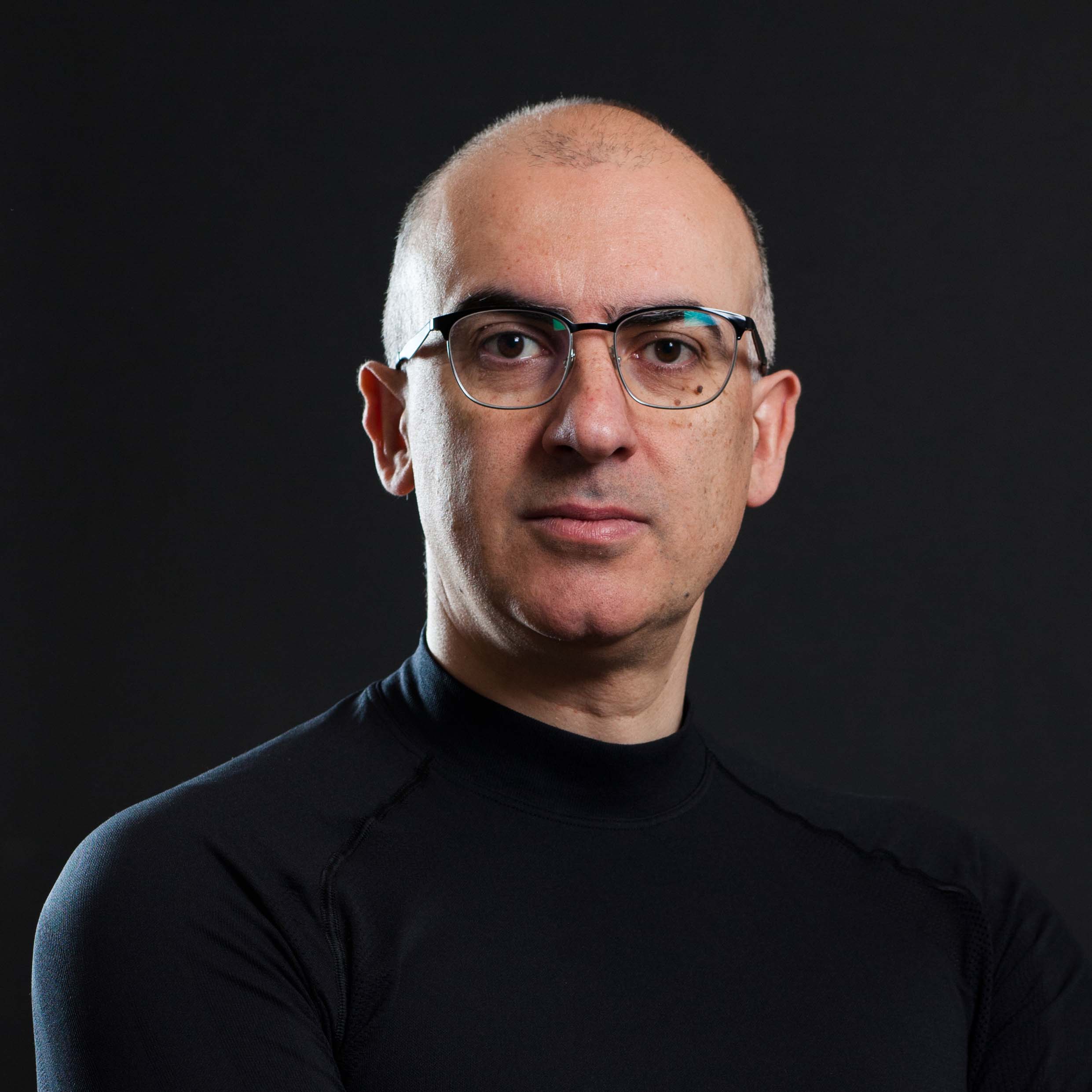}}]{Mario Paolone}
(M’07–SM’10) received the M.Sc. (Hons.) and Ph.D. degrees in electrical engineering from the University of Bologna, Italy, in 1998 and 2002. In 2005, he was an Assistant Professor in power systems with the University of Bologna, where he was with the Power Systems Laboratory until 2011. Since 2011, he has been with the Swiss Federal Institute of Technology, Lausanne, Switzerland, where he is currently Full Professor and the Chair of the Distributed Electrical Systems Laboratory. His research interests focus on power systems with particular reference to real-time monitoring and operation aspects, power system protections, dynamics and transients. Dr. Paolone has authored or co-authored over 300 papers published in mainstream journals and  international conferences in the area of energy and power systems that received numerous awards including the 2013 IEEE EMC Technical Achievement Award, two IEEE Transactions on EMC best paper awards and the Basil Papadias best paper award at the 2013 IEEE PowerTech. Dr. Paolone was the founder Editor-in-Chief of the Elsevier journal Sustainable Energy, Grids and Networks.
\end{IEEEbiography}

\end{document}